\documentclass[runningheads,table,xcdraw]{llncs}

\usepackage[utf8]{inputenc}
\usepackage{amsmath}
\usepackage{microtype}
\usepackage{booktabs}
\usepackage{subcaption}
\usepackage{multirow}
\usepackage{tabularx}
\usepackage[hyphens]{url}
\usepackage{hyperref}
\usepackage{todonotes}
\usepackage{wasysym}
\usepackage{array}
\usepackage{adjustbox}
\usepackage{xstring}
\usepackage{xifthen}

\newcolumntype{x}[1]{>{\centering\arraybackslash}p{#1}}



\usepackage{xcolor}
\usepackage{tikz}

\usepackage[backend=biber,style=lncs]{biblatex}
\usepackage{todonotes}

\usepackage{xspace}

\begin{document}

\title{Logic and Accuracy Testing: A Fifty-State Review}
\author{Josiah Walker \and Nakul Bajaj \and Braden L. Crimmins \and J. Alex Halderman}
\authorrunning{Walker, Bajaj, Crimmins, and Halderman}
\institute{University of Michigan, Ann Arbor, USA\\ \email{\{jhwalker,nbajaj,bradenlc,jhalderm\}@umich.edu}}

\maketitle

\begin{abstract}
    Pre-election logic and accuracy (L\&A) testing is a process in which election officials validate the behavior of voting equipment by casting a known set of test ballots and confirming the expected results. Ideally, such testing can serve to detect certain forms of human error or fraud and help bolster voter confidence. We present the first detailed analysis of L\&A testing practices across the United States. We find that while all states require L\&A testing before every election, their implementations vary dramatically in scope, transparency, and rigorousness. We summarize each state's requirements and score them according to uniform criteria. We also highlight best practices and flag opportunities for improvement, in hopes of encouraging broader adoption of more effective L\&A processes.\looseness=-1
    
\end{abstract}  

\section{Introduction}

The vast majority of votes in the United States are counted mechanically, either by optical scanners that read paper ballots or by direct-recording electronic (DRE) voting machines~\cite{verifier}. To validate that these tabulation devices are configured and functioning correctly, jurisdictions perform a procedure called ``logic and accuracy testing'' (``L\&A testing'') shortly before each election. It typically involves casting a ``test deck''---a set of ballots with known votes---on each machine, then printing the results and ensuring the tally is as expected. Any deviation is a potential indicator that the election equipment has misbehaved.

While more sophisticated mechanisms such as risk-limiting audits~\cite{starksimplerla}, and end-to-end verification~\cite{eevintro} can reliably detect and recover from both errors and attacks after the fact, they are not yet widely applied in the U.S\@. Even if they were, L\&A testing would remain useful for heading off some sources of error before they affected results. Ideally, L\&A testing can protect against certain kinds of malfunction, configuration error, and fraud as well as strengthen voter confidence, but its effectiveness depends on many details of how the testing is performed. In the U.S., L\&A testing requirements---like most aspects of election procedure and the selection of voting equipment---are determined by individual states, resulting in a diversity of practices with widely varying utility.

Unfortunately, this heterogeneity means that many states diverge negatively from the norm and makes it difficult to offer the national public any blanket assurances about the degree of protection that L\&A testing affords. Moreover, many states do not publish detailed L\&A procedures, leaving voters with little ability to assess the effectiveness of their own states' rules, let alone whether any tests they observe comply with them. Yet this decentralized regulatory environment has also allowed a variety of positive L\&A testing procedures to evolve, and there are abundant opportunities for the exchange of best practices.

This paper provides the first comparative analysis of L\&A testing requirements across the fifty states. To determine how each state performs L\&A testing, we conducted an extensive review of available documentation and reached out to election officials in every state. We then assessed and scored each state's policy using criteria designed to reflect its functional effectiveness and suitability as a basis for voter confidence. The results provide a detailed understanding of how states' procedures differ and how well they approach an ideal model of what L\&A testing can achieve. Our analysis reveals that several important L\&A criteria are absent in many or most states' rules, yet we also highlight specific examples of policies that could serve as models for broader dissemination. We hope this work will encourage the adoption of more effective L\&A testing requirements across the United States and help promote policies that better inspire public trust.

\section{Background}

\subsection{L\&A Testing Goals}

L\&A testing was first introduced in the early 1900s for lever-style voting machines~\cite{caliarchive}, which contained a mechanical counter for each candidate. The counters were susceptible to becoming jammed due to physical failure or tampering, so tests were designed to establish that each counter would advance when voted.

Modern DRE voting machines and ballot scanners can suffer from analogous problems---miscalibrated touch-screens or dirty scanner heads can prevent votes in specific ballot positions from being recorded~\cite{nh-sb43-audit}---but they also have more complex failure modes that call for different forms of testing. These devices must be provisioned with an ``election definition'' that specifies the ballot layout and rules. If the election definition is wrong---for instance, the order or position of voting targets do not match the ballots a scanner will read---votes may be miscounted.\looseness=-1

Problems with election definitions caused by human error are surprisingly common. They contributed to the publication of incorrect initial election results in Northampton County, Pennsylvania, in 2019~\cite{northampton-errors}, Antrim County, Michigan, in 2020~\cite{antrim-report}, and DeKalb County, Georgia, in 2022~\cite{dekalb-errors}. In these documented cases the errors were fortunately detected, but only after the results were announced. They likely could have been prevented in the first place by sufficient L\&A testing.

L\&A testing can also serve a role in election security. Research has long recognized that L\&A testing \emph{cannot} reliably defeat an adversary who manages to execute malware on voting machines, because the malware could detect when it was under test and only begin cheating during the election itself (see, e.g.,~\cite{feldman07}). However, L\&A testing can potentially thwart more limited attackers who manage to tamper with election definitions or configuration settings. For example, although there is no evidence that the instances of error described above were caused by fraud, attackers could cause similar election definition problems deliberately in an attempt to alter results. This would likely require far less sophistication than creating vote-stealing malware. Moreover, there is growing concern about threats posed by dishonest election insiders, who routinely have the access necessary to perform such an attack~\cite{inside-threats-washpo}.

Beyond providing these protections, L\&A testing also frequently serves a role in enhancing public confidence in elections. Most states conduct at least part of their L\&A testing during a public ceremony, where interested political party representatives, candidates, news media, and residents can observe the process and sometimes even participate by marking test ballots. Some jurisdictions also provide live or recorded video of their testing ceremonies online. These public tests can help build trust by allowing voters to meet their local officials, observe their level of diligence, and become more familiar with election processes. Additionally, public observers have the potential to make testing stronger, by providing an independent check that the required tests were completed and performed correctly. At least in principle, public observation could also help thwart attempts by dishonest officials to subvert L\&A testing by skipping tests or ignoring errors.\looseness=-1

\subsection{U.S. Elections}

L\&A testing fills a role that is best understood with a view towards the broader context of election administration in the jurisdictions where it is practiced. In the U.S., many subjects are put to the voters, frequently all at once, and a single ballot might include contests ranging from the national presidency and congress to the state governor,  legislature, and judges to the local mayor, city council, sheriff, and school board~\cite{sample-ballot}. This means elections tend to involve many contests---typically around 20, although some jurisdictions have occasionally had nearly 100~\cite{ballot-lengths}. There may also be several ballot variants within a single polling place to accommodate candidates from different sets of districts. These features make tallying by hand impracticable in many areas. As a result, nearly all jurisdictions rely on electronic tabulation equipment, today most commonly in the form of computerized ballot scanners~\cite{verifier}. Ensuring that these machines are properly configured and functioning on election day is the key motivation for L\&A testing.\looseness=-1

Election administration in the U.S. is largely the province of state and local governments. Although the Constitution gives Congress the power to override state law regarding the ``manner of holding Elections for Senators and Representatives,'' this authority has been applied only sparingly, for instance to establish accessibility requirements and enforce civil rights~\cite{hava-2002, vra-1965}. Each state legislature establishes its own election laws, and the state executive (typically the secretary of state) promulgates more detailed regulations and procedures. In practice, election administration powers are exercised primarily by local jurisdictions, such as counties or cities and townships, where local officials (often elected officials called ``clerks'') are responsible for conducting elections~\cite{ncsl-local-admin}.

Because of this structure, there is little standardization of election practices across the states, and L\&A testing is no exception. Testing processes (and the ceremonies that accompany them) vary substantially between and within states. As we show, these variations have significant effects, both with respect to error-detection effectiveness and procedural transparency and intelligibility. Pessimistically, one can view this broad local discretion as a way for some jurisdictions to use lax practices with little accountability. We note, however, that it also grants many clerks the power to depart \emph{upwards} from their states'  mandatory procedures, achieving stronger protections than the law requires. This provides an opportunity for improved practices to see early and rapid adoption.

\subsection{Related Work}

Although L\&A testing itself has so far received little research attention, there is extensive literature analyzing other aspects of election mechanics across states and countries, with the goal of informing policymaking and spreading best practices. For instance, past work has examined state practices and their impacts regarding post-election audits~\cite{eac-election-audits}, voter registration list maintenance~\cite{caltech-voter-reg-lists}, voter identification requirements~\cite{cambridge-voter-id}, online voter registration~\cite{online-voter-reg-turnout}, election observation laws~\cite{election-observation-laws}, the availability of universal vote-by-mail~\cite{universal-vbm}. A far larger body of research exists comparing state practices in fields other than elections.

Despite the abundance of this work, we are the first (to our knowledge) to examine states' L\&A testing practices in detail. A 2018 state-by-state report by the Center for American Progress~\cite{CAP-report-2018} considered L\&A testing among several other aspects of election security preparedness; however, it primarily focused  on the narrow question of whether states required all equipment to be tested. To build upon this research, we consider many other policy choices that influence the effectiveness of L\&A requirements and procedures. 

\section{Methodology}

\subsection{Data Collection}

To gather information on states' practices, we began by collecting official documentation where publicly available, relying primarily on state legal codes, state election websites, and Internet search engines. If we could not locate sufficient information, we attempted to contact the state via email or by phone to supplement our understanding or ask for clarifications. We directed these inquiries to the state elections division's main contact point, as identified on its website.

State responses varied. While some states provided line by line answers to each of our questions, it was common for states to indicate that our criteria were more specific than what state resources dictated, pointing us instead to the same statues and documentation we had already examined, providing us with additional documentation that was still unresponsive to our questions, or replying in paragraphs that partially addressed some questions while completely disregarding others. In cases where we could not find evidence to support that a state satisfied certain criteria and the state did not provide supporting evidence upon request, we did not award the state any points for those criteria.

Upon finalizing our summary of each state's practices, we contacted officials again to provide an opportunity for them to complete or correct our understanding. Over the course of nine months, we communicated with all 50 states and received at least some feedback on our summaries from all but seven states---Iowa, New Jersey, New York, Rhode Island, Tennessee, Vermont, and Wisconsin. Our data and analysis are current as of July 2022.

\subsection{Evaluation Criteria}

To uniformly assess and compare states practices, we applied the following criteria and scoring methodology, which reflect attributes we consider important for maximizing the benefits of L\&A testing in terms of accuracy and voter confidence. These criteria are non-exhaustive, but we believe they are sufficiently comprehensive to evaluate state procedures relative to one another. (Additional desirable testing properties are discussed in Section~\ref{sec:discussion}.) Note that our assessments do not necessarily reflect practice in each of a state's subdivisions, since local officials sometimes have authority to exceed state guidelines. To keep the analysis tractable, we instead focus on the \emph{baseline} established by statewide requirements.

We developed two categories of criteria: \emph{procedural criteria}, which encompass the existence of procedures, the scope of testing, and transparency; and \emph{functional criteria}, which reflect whether the testing could reliably detect various kinds of errors and attacks. To facilitate quantitative comparisons, we assigned point values to each criterion, such that each category is worth a total of 10 points and the weights of specific items reflect our assessment of their relative importance.

\medskip
\noindent{\bfseries Procedural Criteria}

\smallskip
\noindent{\bfseries\em Rules and Transparency} (5 points)\quad
To provide the strongest basis for trust, testing should meet or exceed published requirements and be conducted in public.\vspace{-6pt}

\begin{description}
\setlength{\itemsep}{\smallskipamount}
\item RT1 ($1.5$ pts): Procedures are specified in a detailed public document.\smallskip

This captures the threshold matter of whether states have published L\&A requirements. Detailed or step-by-step guidelines received full credit, and general laws or policies received half credit.

\item RT2 ($1.0$ pts): The document is readily available, e.g., via the state's website.\smallskip

Making L\&A procedures easily available helps inform the public and enables observers to assess tests they witness.\footnote{Even when procedures are public documents, they are not always readily accessible. One state, Delaware, instructed us that we would need to find a resident to file a Freedom of Information Act request before their procedures would be provided.}

\item RT3 ($1.5$ pts): Some testing is open to the public, candidates/parties, journalists.\smallskip

This tracks the potential for public L\&A ceremonies to strengthen confidence.

\item RT4 ($1.0$ pts): Local jurisdictions have latitude to exceed baseline requirements.
\end{description}
\vspace{-6pt}

\noindent{\bfseries\em Scope of Testing} (5 points)\quad
A comprehensive approach to testing covers every ballot design across all the voting machines or scanners where they can be used.\vspace{-6pt}

\begin{description}
\setlength{\itemsep}{\smallskipamount}
\item ST1 ($2.0$ pts): All voting machines/scanners must be tested before each election.
\item ST2 ($1.0$ pts): All devices must be tested \emph{at a public event} before each election.
\item ST3 ($2.0$ pts): All devices must be tested with every applicable ballot design.
\smallskip

Failing to test all machines or all ballot styles risks that localized problems will go undetected, so each was assigned a substantial 2 points. One additional point was provided if all testing is public, to reflect transparency interests.
\end{description}
\vspace{-8pt}

\medskip
\noindent{\bfseries Functional Criteria}

\smallskip
\noindent In each of three sets of functional criteria, we assess a simple form of the protection (with a small point value) and a more rigorous form (with a large point value).

\smallskip
\noindent{\bfseries\em Basic Protections} (4 points)\quad
To guard against common errors, tests should cover every voting target and ensure detection of transpositions.\vspace{-6pt}

\begin{description}
\setlength{\itemsep}{\smallskipamount}
\item BP1 ($1.0$ pts): All choices receive at least one valid vote during testing.
\item BP2 ($3.0$ pts): No two choices in a contest receive the same number of votes.\smallskip

The first test minimally detects whether each candidate has some functioning voting target. The second further ensures the detection of transposed targets within a contest, which can result from misconfigured election definitions.
\end{description}
\vspace{-6pt}

\smallskip
\noindent{\bfseries\em Overvote Protection} (2 points)\quad
Testing should exercise overvote detection and, ideally, confirm that the overvote threshold in each contest is set correctly.\vspace{-6pt}

\begin{description}
\setlength{\itemsep}{\smallskipamount}
\item OP1 ($0.5$ pts): At least one overvoted ballot is cast during testing.
\item OP2 ($1.5$ pts): For each contest $c$, a test deck includes a ballot with $n_c$ selections and one with $n_c+1$ selections, where $n_c$ is the permitted number of selections.\smallskip

An overvote occurs when the voter selects more than the permitted number of candidates, rendering the selections invalid. The first practice minimally detects that the machine is configured to reject overvotes, while the second tests that the allowed number of selections is set correctly for each contest.
\end{description}
\vspace{-6pt}

\smallskip
\noindent{\bfseries\em Nondeterministic Testing} (4 points)\quad
For stronger protection against deliberate errors, attackers should be unable to predict how the test deck is marked.\vspace{-6pt}

\begin{description}
\setlength{\itemsep}{\smallskipamount}
\item ND1 ($1.0$ pts): Public observers are allowed to arbitrarily mark and cast ballots.
\item ND2 ($3.0$ pts): Some ballots must be marked using a source of randomness.\smallskip

Attackers who can predict the test deck potentially can tamper with the election definition such that errors will not be visible during testing. If the public can contribute test ballots, this introduces uncertainty for the attacker, while requiring random ballots allows for more rigorous probabilistic detection.\looseness=-1

\end{description}
\vspace{-6pt}

\newcommand{\symbyes}{\CIRCLE}
\newcommand{\symbhalf}{\LEFTcircle}
\newcommand{\symbno}{\Circle}
\newcommand{\symbunk}{$\odot$}

\begin{table}
\begin{adjustbox}{width=\linewidth}
\centering
\begin{tabular}{ccccclcccrlccclcclccr}
\cmidrule[\heavyrulewidth]{1-10}\cmidrule[\heavyrulewidth]{12-21}
\multirow{3.4}{*}{\rotatebox[origin=c]{270}{State}}
& \multicolumn{9}{c}{\bf Procedural Criteria} & & 
\multirow{3.4}{*}{\rotatebox[origin=c]{270}{State}}
& \multicolumn{9}{c}{\bf Functional Criteria} \\
\cmidrule{2-10}\cmidrule{13-21}

 & RT1 & RT2 & RT3 & RT4 & \hspace{5pt} & ST1 & ST2 & ST3 & \multicolumn{1}{c}{$\sum$} & \strut\hspace{0.75cm}\strut &
 & BP1 & BP2 & \hspace{5pt} & OP1 & OP2 & \hspace{5pt} & ND1 & ND2 & \multicolumn{1}{c}{$\sum$} \\
 & 1.5 & 1.0 & 1.5 & 1.0 & & 2.0 & 1.0 & 2.0 & \multicolumn{1}{c}{(/10)} & & 
 & 1.0 & 3.0 & & 0.5 & 1.5 & & 1.0 & 3.0 & \multicolumn{1}{c}{(/10)} \\

\cmidrule[\lightrulewidth]{1-10}\cmidrule[\lightrulewidth]{12-21}

MT & \symbyes & \symbyes & \symbyes & \symbyes &  & \symbyes & \symbyes & \symbyes & 10.00 &  & CT & \symbyes & \symbyes &  & \symbyes & \symbyes &  & \symbno & \symbhalf & 7.50 \\OH & \symbyes & \symbyes & \symbyes & \symbyes &  & \symbyes & \symbyes & \symbyes & 10.00 &  & AZ & \symbyes & \symbunk &  & \symbyes & \symbyes &  & \symbyes & \symbyes & 7.00 \\PA & \symbyes & \symbyes & \symbyes & \symbyes &  & \symbyes & \symbyes & \symbyes & 10.00 &  & SD & \symbyes & \symbyes &  & \symbyes & \symbyes &  & \symbyes & \symbno & 7.00 \\WA & \symbyes & \symbyes & \symbyes & \symbyes &  & \symbyes & \symbyes & \symbyes & 10.00 &  & IL & \symbyes & \symbyes &  & \symbyes & \symbyes &  & \symbno & \symbno & 6.00 \\CO & \symbhalf & \symbyes & \symbyes & \symbyes &  & \symbyes & \symbyes & \symbyes & 9.25 &  & MO & \symbyes & \symbyes &  & \symbyes & \symbyes &  & \symbno & \symbno & 6.00 \\CT & \symbhalf & \symbyes & \symbyes & \symbyes &  & \symbyes & \symbyes & \symbyes & 9.25 &  & NE & \symbyes & \symbyes &  & \symbyes & \symbyes &  & \symbunk & \symbunk & 6.00 \\MA & \symbhalf & \symbyes & \symbyes & \symbyes &  & \symbyes & \symbyes & \symbyes & 9.25 &  & MI & \symbyes & \symbyes &  & \symbyes & \symbno &  & \symbyes & \symbno & 5.50 \\MO & \symbhalf & \symbyes & \symbyes & \symbyes &  & \symbyes & \symbyes & \symbyes & 9.25 &  & UT & \symbyes & \symbyes &  & \symbyes & \symbno &  & \symbyes & \symbno & 5.50 \\NV & \symbhalf & \symbyes & \symbyes & \symbyes &  & \symbyes & \symbyes & \symbyes & 9.25 &  & VT & \symbyes & \symbyes &  & \symbunk & \symbunk &  & \symbunk & \symbhalf & 5.50 \\SD & \symbhalf & \symbyes & \symbyes & \symbyes &  & \symbyes & \symbyes & \symbyes & 9.25 &  & WY & \symbyes & \symbyes &  & \symbyes & \symbno &  & \symbyes & \symbno & 5.50 \\UT & \symbhalf & \symbyes & \symbyes & \symbyes &  & \symbyes & \symbyes & \symbyes & 9.25 &  & AR & \symbyes & \symbyes &  & \symbyes & \symbunk &  & \symbunk & \symbunk & 4.50 \\VA & \symbyes & \symbyes & \symbhalf & \symbyes &  & \symbyes & \symbyes & \symbyes & 9.25 &  & DE & \symbyes & \symbyes &  & \symbyes & \symbno &  & \symbno & \symbno & 4.50 \\WV & \symbhalf & \symbyes & \symbyes & \symbyes &  & \symbyes & \symbyes & \symbyes & 9.25 &  & FL & \symbyes & \symbyes &  & \symbyes & \symbunk &  & \symbunk & \symbno & 4.50 \\WI & \symbhalf & \symbyes & \symbyes & \symbyes &  & \symbyes & \symbyes & \symbyes & 9.25 &  & MN & \symbyes & \symbyes &  & \symbyes & \symbno &  & \symbunk & \symbunk & 4.50 \\WY & \symbhalf & \symbyes & \symbyes & \symbyes &  & \symbyes & \symbyes & \symbyes & 9.25 &  & MT & \symbyes & \symbyes &  & \symbyes & \symbno &  & \symbno & \symbno & 4.50 \\MI & \symbyes & \symbyes & \symbyes & \symbyes &  & \symbyes & \symbno & \symbyes & 9.00 &  & ND & \symbyes & \symbyes &  & \symbyes & \symbno &  & \symbno & \symbno & 4.50 \\MN & \symbyes & \symbyes & \symbyes & \symbyes &  & \symbyes & \symbno & \symbyes & 9.00 &  & OH & \symbyes & \symbyes &  & \symbyes & \symbunk &  & \symbunk & \symbunk & 4.50 \\NH & \symbyes & \symbyes & \symbyes & \symbunk &  & \symbyes & \symbyes & \symbyes & 9.00 &  & WA & \symbyes & \symbyes &  & \symbyes & \symbno &  & \symbno & \symbno & 4.50 \\VT & \symbyes & \symbyes & \symbyes & \symbunk &  & \symbyes & \symbyes & \symbyes & 9.00 &  & ID & \symbyes & \symbno &  & \symbyes & \symbyes &  & \symbyes & \symbno & 4.00 \\ID & \symbhalf & \symbhalf & \symbyes & \symbyes &  & \symbyes & \symbyes & \symbyes & 8.75 &  & NM & \symbyes & \symbno &  & \symbyes & \symbyes &  & \symbno & \symbno & 3.00 \\AK & \symbyes & \symbhalf & \symbyes & \symbunk &  & \symbyes & \symbyes & \symbyes & 8.50 &  & CO & \symbyes & \symbno &  & \symbyes & \symbno &  & \symbyes & \symbno & 2.50 \\DE & \symbyes & \symbhalf & \symbyes & \symbunk &  & \symbyes & \symbyes & \symbyes & 8.50 &  & IA & \symbyes & \symbunk &  & \symbyes & \symbunk &  & \symbyes & \symbunk & 2.50 \\GA & \symbyes & \symbhalf & \symbyes & \symbunk &  & \symbyes & \symbyes & \symbyes & 8.50 &  & AL & \symbyes & \symbunk &  & \symbyes & \symbunk &  & \symbunk & \symbunk & 1.50 \\NC & \symbyes & \symbhalf & \symbyes & \symbunk &  & \symbyes & \symbyes & \symbyes & 8.50 &  & GA & \symbyes & \symbunk &  & \symbyes & \symbunk &  & \symbno & \symbunk & 1.50 \\AL & \symbhalf & \symbyes & \symbyes & \symbyes &  & \symbyes & \symbno & \symbyes & 8.25 &  & IN & \symbyes & \symbunk &  & \symbyes & \symbunk &  & \symbno & \symbno & 1.50 \\ND & \symbhalf & \symbyes & \symbyes & \symbunk &  & \symbyes & \symbyes & \symbyes & 8.25 &  & KS & \symbyes & \symbunk &  & \symbyes & \symbunk &  & \symbunk & \symbunk & 1.50 \\AZ & \symbyes & \symbyes & \symbyes & \symbyes &  & \symbyes & \symbyes & \symbunk & 8.00 &  & KY & \symbyes & \symbunk &  & \symbyes & \symbunk &  & \symbunk & \symbunk & 1.50 \\IL & \symbyes & \symbhalf & \symbyes & \symbyes &  & \symbyes & \symbyes & \symbno & 7.50 &  & ME & \symbyes & \symbunk &  & \symbyes & \symbunk &  & \symbno & \symbunk & 1.50 \\FL & \symbyes & \symbyes & \symbyes & \symbyes &  & \symbyes & \symbno & \symbno & 7.00 &  & NV & \symbyes & \symbunk &  & \symbyes & \symbunk &  & \symbunk & \symbunk & 1.50 \\IN & \symbyes & \symbyes & \symbyes & \symbyes &  & \symbno & \symbno & \symbyes & 7.00 &  & NH & \symbyes & \symbunk &  & \symbyes & \symbunk &  & \symbunk & \symbunk & 1.50 \\SC & \symbyes & \symbhalf & \symbyes & \symbunk &  & \symbyes & \symbyes & \symbunk & 6.50 &  & NJ & \symbyes & \symbunk &  & \symbyes & \symbunk &  & \symbno & \symbunk & 1.50 \\IA & \symbhalf & \symbyes & \symbyes & \symbunk &  & \symbyes & \symbyes & \symbunk & 6.25 &  & NY & \symbyes & \symbunk &  & \symbyes & \symbunk &  & \symbunk & \symbunk & 1.50 \\KS & \symbhalf & \symbyes & \symbyes & \symbunk &  & \symbyes & \symbyes & \symbunk & 6.25 &  & PA & \symbyes & \symbno &  & \symbyes & \symbunk &  & \symbno & \symbunk & 1.50 \\MS & \symbhalf & \symbyes & \symbyes & \symbunk &  & \symbyes & \symbyes & \symbunk & 6.25 &  & TN & \symbyes & \symbunk &  & \symbyes & \symbunk &  & \symbunk & \symbunk & 1.50 \\RI & \symbhalf & \symbyes & \symbyes & \symbunk &  & \symbyes & \symbyes & \symbunk & 6.25 &  & TX & \symbyes & \symbno &  & \symbyes & \symbno &  & \symbno & \symbno & 1.50 \\TX & \symbhalf & \symbyes & \symbyes & \symbyes &  & \symbno & \symbno & \symbyes & 6.25 &  & VA & \symbyes & \symbunk &  & \symbyes & \symbunk &  & \symbunk & \symbunk & 1.50 \\AR & \symbyes & \symbyes & \symbyes & \symbunk &  & \symbyes & \symbunk & \symbunk & 6.00 &  & WV & \symbyes & \symbunk &  & \symbyes & \symbunk &  & \symbunk & \symbunk & 1.50 \\NE & \symbyes & \symbhalf & \symbunk & \symbunk &  & \symbyes & \symbunk & \symbyes & 6.00 &  & WI & \symbyes & \symbunk &  & \symbyes & \symbunk &  & \symbunk & \symbno & 1.50 \\KY & \symbhalf & \symbyes & \symbno & \symbunk &  & \symbyes & \symbno & \symbyes & 5.75 &  & HI & \symbno & \symbno &  & \symbyes & \symbno &  & \symbhalf & \symbno & 1.00 \\MD & \symbhalf & \symbyes & \symbno & \symbunk &  & \symbyes & \symbno & \symbyes & 5.75 &  & MD & \symbyes & \symbno &  & \symbno & \symbno &  & \symbno & \symbno & 1.00 \\NY & \symbhalf & \symbyes & \symbunk & \symbunk &  & \symbyes & \symbunk & \symbyes & 5.75 &  & SC & \symbyes & \symbunk &  & \symbno & \symbunk &  & \symbunk & \symbunk & 1.00 \\CA & \symbhalf & \symbyes & \symbhalf & \symbyes &  & \symbyes & \symbno & \symbno & 5.50 &  & MA & \symbno & \symbno &  & \symbyes & \symbno &  & \symbno & \symbno & 0.50 \\LA & \symbhalf & \symbyes & \symbhalf & \symbunk &  & \symbyes & \symbyes & \symbunk & 5.50 &  & MS & \symbunk & \symbunk &  & \symbyes & \symbunk &  & \symbno & \symbunk & 0.50 \\NM & \symbhalf & \symbyes & \symbhalf & \symbno &  & \symbyes & \symbyes & \symbno & 5.50 &  & NC & \symbunk & \symbunk &  & \symbyes & \symbunk &  & \symbno & \symbunk & 0.50 \\HI & \symbyes & \symbhalf & \symbno & \symbyes &  & \symbyes & \symbno & \symbno & 5.00 &  & OK & \symbunk & \symbunk &  & \symbyes & \symbunk &  & \symbno & \symbunk & 0.50 \\OR & \symbhalf & \symbyes & \symbhalf & \symbunk &  & \symbyes & \symbno & \symbunk & 4.50 &  & AK & \symbunk & \symbunk &  & \symbunk & \symbunk &  & \symbunk & \symbunk & 0.00 \\ME & \symbhalf & \symbyes & \symbyes & \symbunk &  & \symbunk & \symbunk & \symbunk & 3.25 &  & CA & \symbno & \symbno &  & \symbno & \symbno &  & \symbno & \symbno & 0.00 \\NJ & \symbhalf & \symbyes & \symbyes & \symbunk &  & \symbunk & \symbunk & \symbunk & 3.25 &  & LA & \symbunk & \symbunk &  & \symbunk & \symbunk &  & \symbno & \symbunk & 0.00 \\OK & \symbyes & \symbhalf & \symbhalf & \symbunk &  & \symbunk & \symbunk & \symbunk & 2.75 &  & OR & \symbunk & \symbunk &  & \symbunk & \symbunk &  & \symbno & \symbunk & 0.00 \\TN & \symbhalf & \symbyes & \symbhalf & \symbunk &  & \symbno & \symbno & \symbunk & 2.50 &  & RI & \symbunk & \symbunk &  & \symbunk & \symbunk &  & \symbunk & \symbunk & 0.00 \\

\cmidrule[\heavyrulewidth]{1-10}\cmidrule[\heavyrulewidth]{12-21}
\end{tabular}
\end{adjustbox}

\smallskip
\hspace{0.5em}{\symbyes: Fully met \hfill \symbhalf: Partly met \hfill \symbno: Not met \hfill \symbunk: State not responsive (scored as unmet)}
\label{tab:summary}
\end{table}

\section{Analysis}

\begin{figure}[b]
    \centering
    \begin{subfigure}{0.5\linewidth}
        \includegraphics[width=\linewidth]{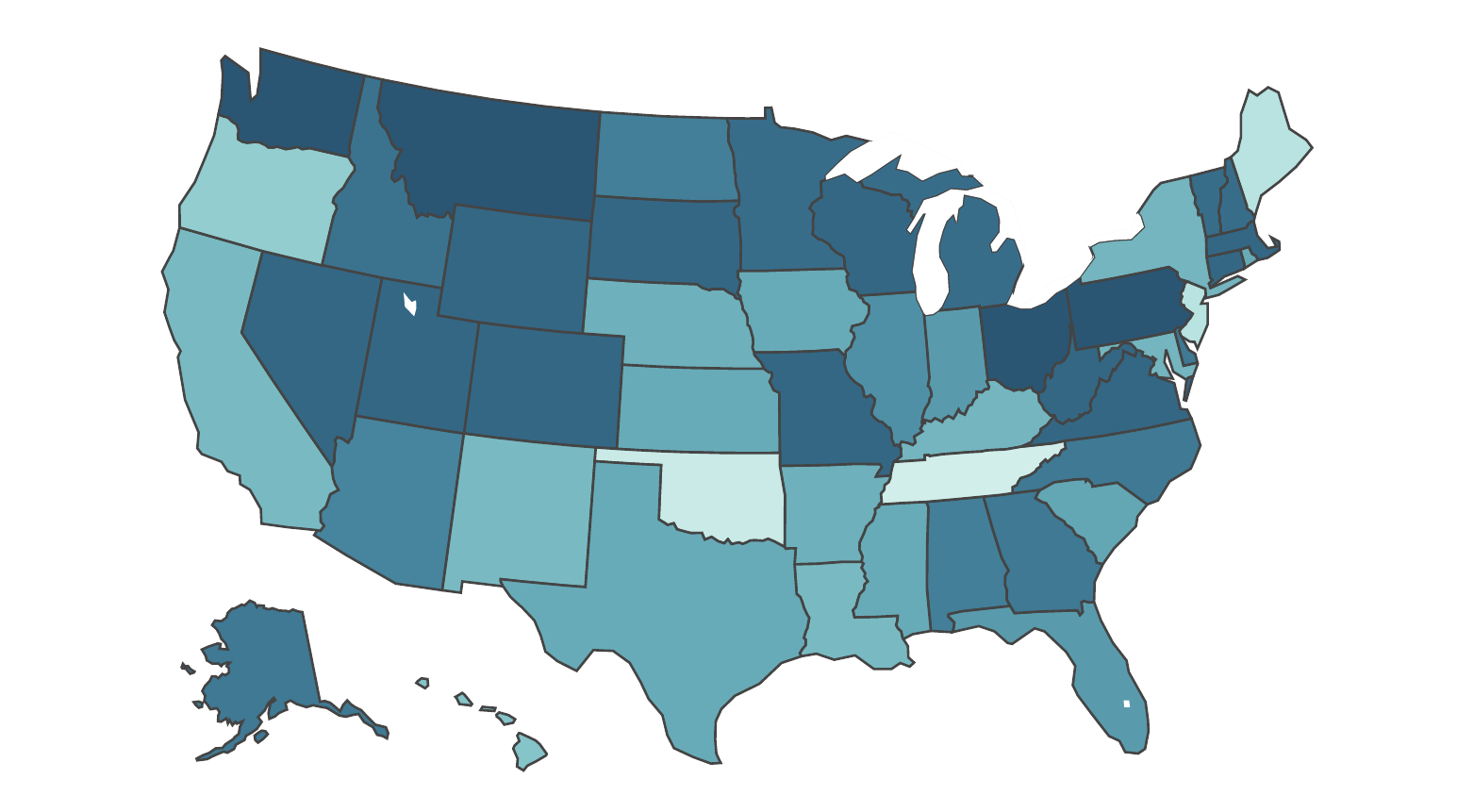}%
        \caption{Procedural Score}
    \end{subfigure}%
    \begin{subfigure}{0.5\linewidth}
        \includegraphics[width=\linewidth]{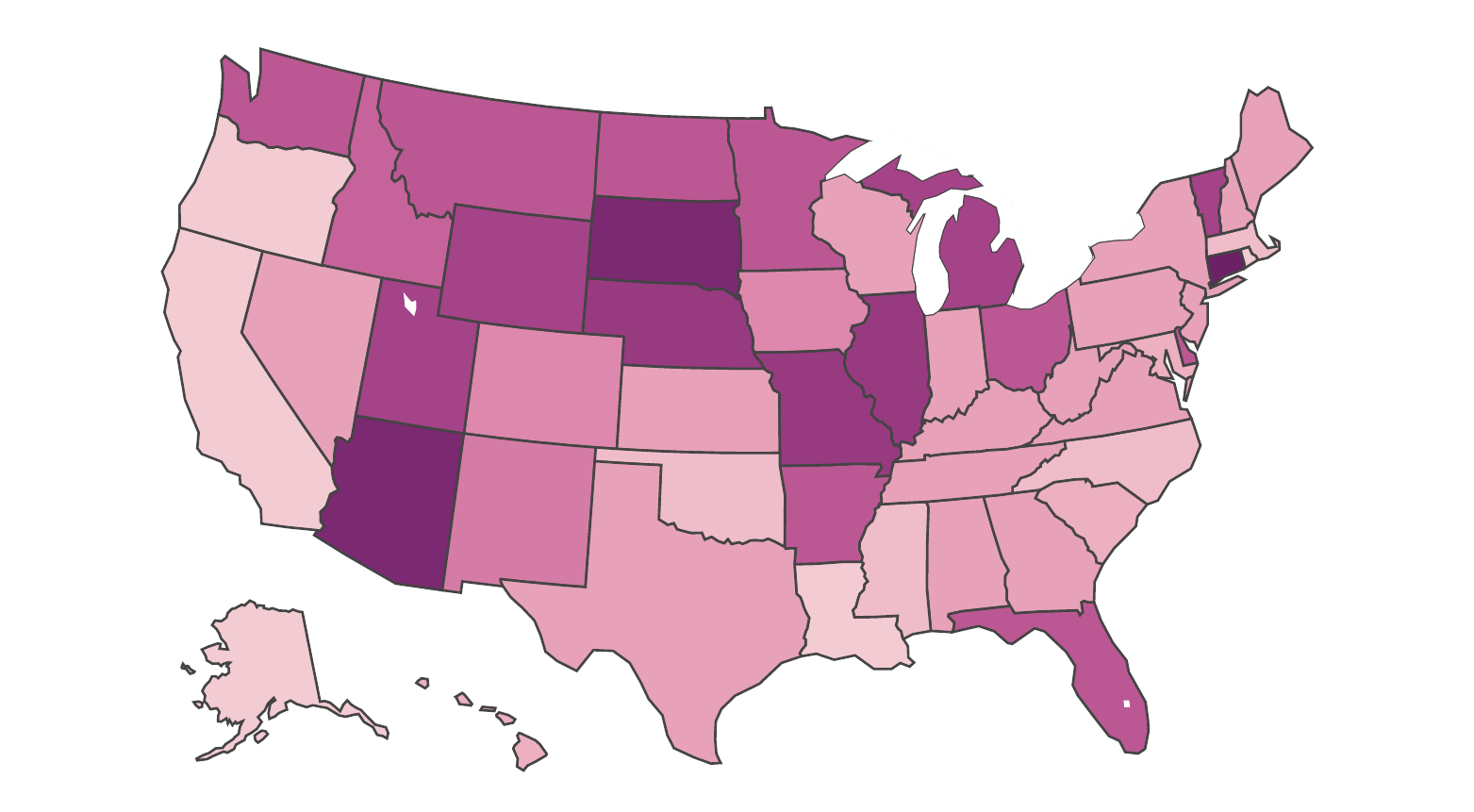}
        \caption{Functional Score}
    \end{subfigure}
    \caption{Mapping state scores (darker indicates better performance) shows that L\&A testing practices vary significantly within all geographic regions of the U.S.}
    \label{fig:maps}
\end{figure}

Our nationwide review of L\&A procedures highlights significant variation among the testing practices of the fifty states, as illustrated by the maps in Figure~\ref{fig:maps}. The tables on page \pageref{tab:summary} summarize our findings and rank the states with respect to the procedural and functional criteria. We also provide a capsule summary of each state's practices in Appendix~\ref{sec:states}.

\begin{figure}[t]
    \centering
    \includegraphics[width=\linewidth]{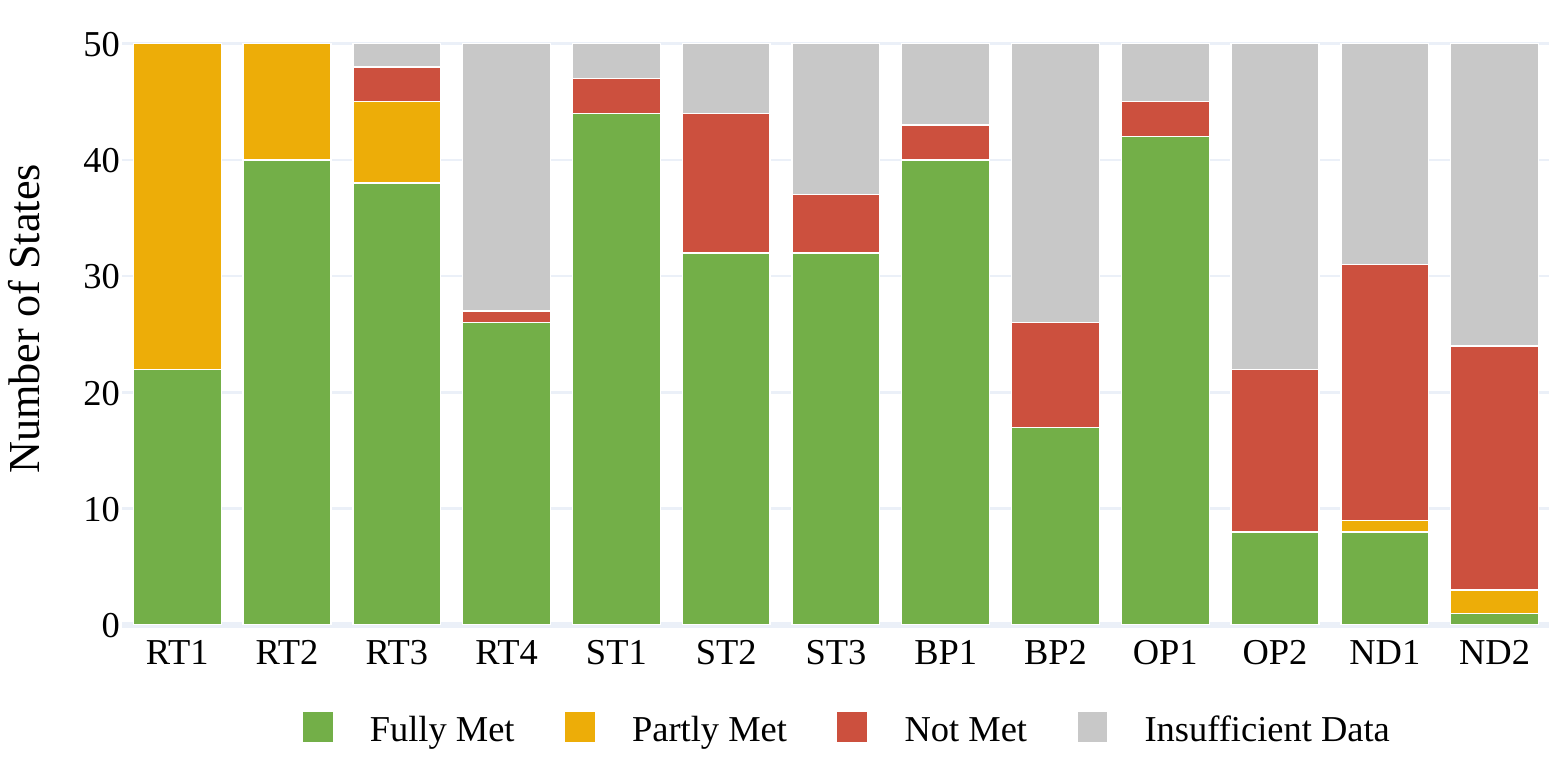}
    \caption{We count the number of states that met, partly met, or did not meet each criterion. While states commonly require simple protections (BP1, OP1), most do not achieve more rigorous forms of error detection (e.g., OP2, ND1, and ND2).\looseness=-1}
    \label{fig:counts}
\end{figure}

\subsection{Performance by Criterion}

Figure~\ref{fig:counts} shows the number of states that met, partly met, or did not meet each criterion. All 50 states have laws that require L\&A testing, but only 22 have a public, statewide document that details the steps necessary to properly conduct the tests (RT1). Of those that do not, several (such as California) merely instruct local jurisdictions to follow instructions from their voting equipment vendors, limiting the efficacy of logic and accuracy procedures to each vendor's preferences. 

States generally performed well with respect to transparency criteria. Every state has some public documentation about its L\&A practices, with 40 states making this documentation readily available (RT2). At least 45 states perform some or all testing in public (RT3), although 7 of these impose restrictions on who is allowed to attend. At least 32 states test every machine in public (ST2). Just three states (Kentucky, Maryland, and Hawaii) do not conduct any public L\&A testing, which may be a significant lost opportunity to build public trust.

Most states also scored high marks regarding the scope of their testing. We were able to confirm that at least 44 states require all equipment to be tested before each election (ST1). Exceptions include Tennessee, Texas, and Indiana, which only require testing a sample of machines. At least 32 states require every ballot style to be tested, but 5 or more do not (ST3), which increases the chances that problems will go undetected in these jurisdictions. 
 
Consideration of several other criteria is more complicated, because we often lack evidence for or against their being met. We have insufficient data about 19 or more states for RT4 and most of the functional criteria (BP2, OP2, ND1, and ND2). Details concerning functional criteria  tended to be less frequently described in public documentation, which potentially biases our analysis in cases where states were also unresponsive to inquiries. We treated such instances as unmet for scoring purposes, but it is also informative to consider the ratio of met to unmet, as depicted in Figure~\ref{fig:counts}. One example is whether states allow local jurisdictions to exceed their baseline requirements (RT4). Although we lack data for 23 states, the criterion is met by at least 26 states, and we have only confirmed that it is unmet in one state (New Mexico). This suggests that many of the unconfirmed states likely also allow local officials to depart upwards from their requirements.\looseness=-1

After accounting for what data was available, states clearly perform much better on our procedural criteria than on our functional criteria. This suggests that many of the functional attributes we looked for are aspirational relative to current practice and indicates that L\&A testing \emph{could} provide much more value.

The only two functional criteria that most states meet are basic protections for voting targets (BP1) and overvotes (OP1), which are provided in at least 40 and 42 states, respectively. At least 17 states would detect transposed voting targets within contests (BP2), but as few as 8 fully validate overvote thresholds (OP2). These more rigorous protections require more complicated procedures and larger test decks, but that some states achieve them suggests that they would be practical to implement more broadly.

Policies facilitating even the basic form of nondeterministic testing were rare. Only 11 states scored even partial points for conducting nondeterministic testing, with 9 of them allowing public observers to mark test ballots (ND1) and 3 of them (Arizona, Connecticut, Vermont) confirming that election officials are required to mark random selections (ND2). Of the three, only Arizona confirmed that it required officials to use a random number generator, thus earning full points. These findings are surprising, since unpredictable or randomized testing can thwart certain kinds of attacks that predictable tests cannot. That nondeterministic testing is rare greatly limits the security benefits of typical state L\&A practices.

\begin{figure}[t]
    \centering
    \includegraphics[width=\linewidth]{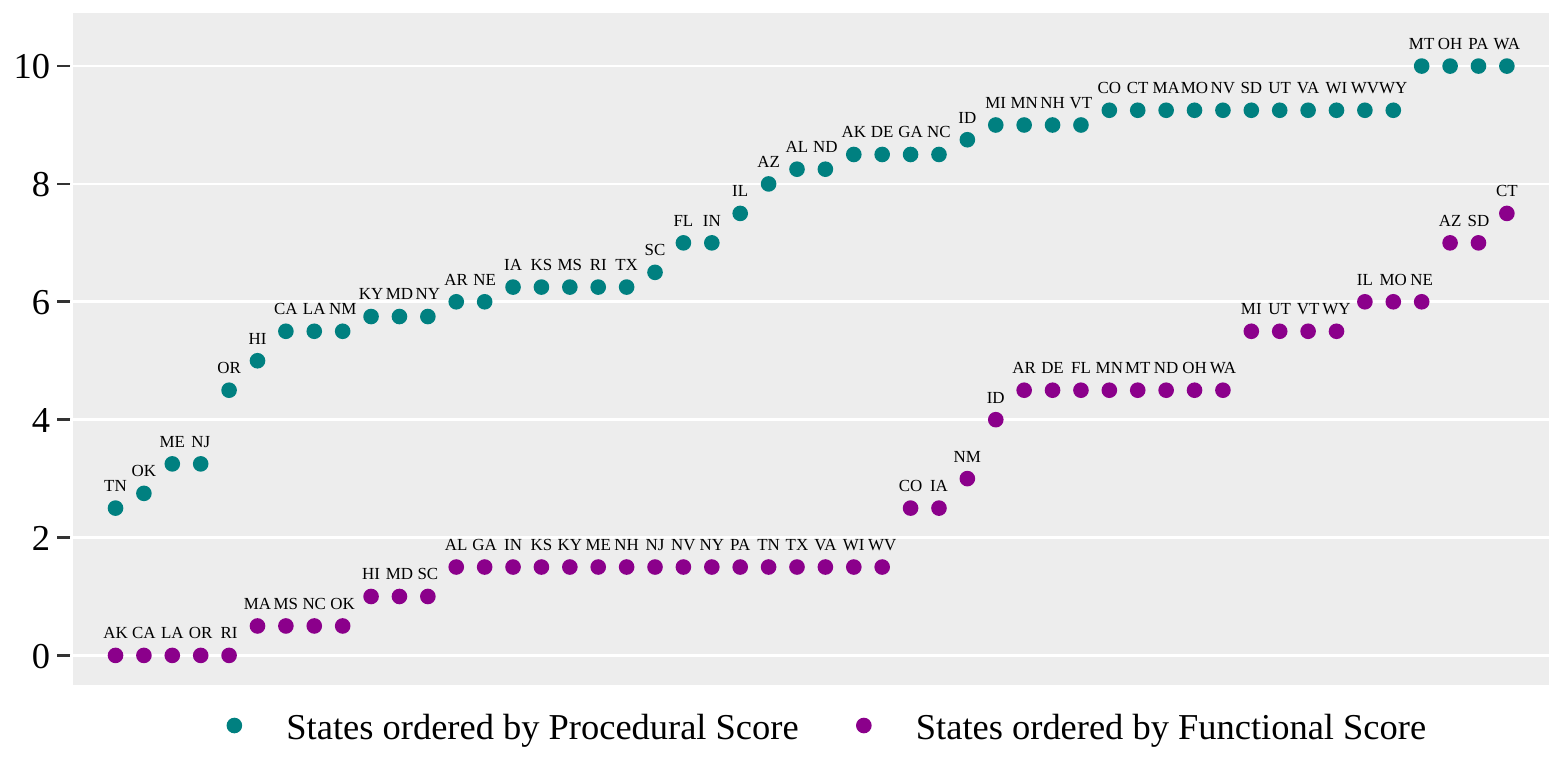}
    \caption{Many states have perfect or near-perfect procedural scores, but functional scores are generally lower, reflecting opportunities for making L\&A more effective.}
    \label{fig:ordered}
\end{figure}
\begin{figure}[t]
    \centering
    \includegraphics[width=0.66\linewidth]{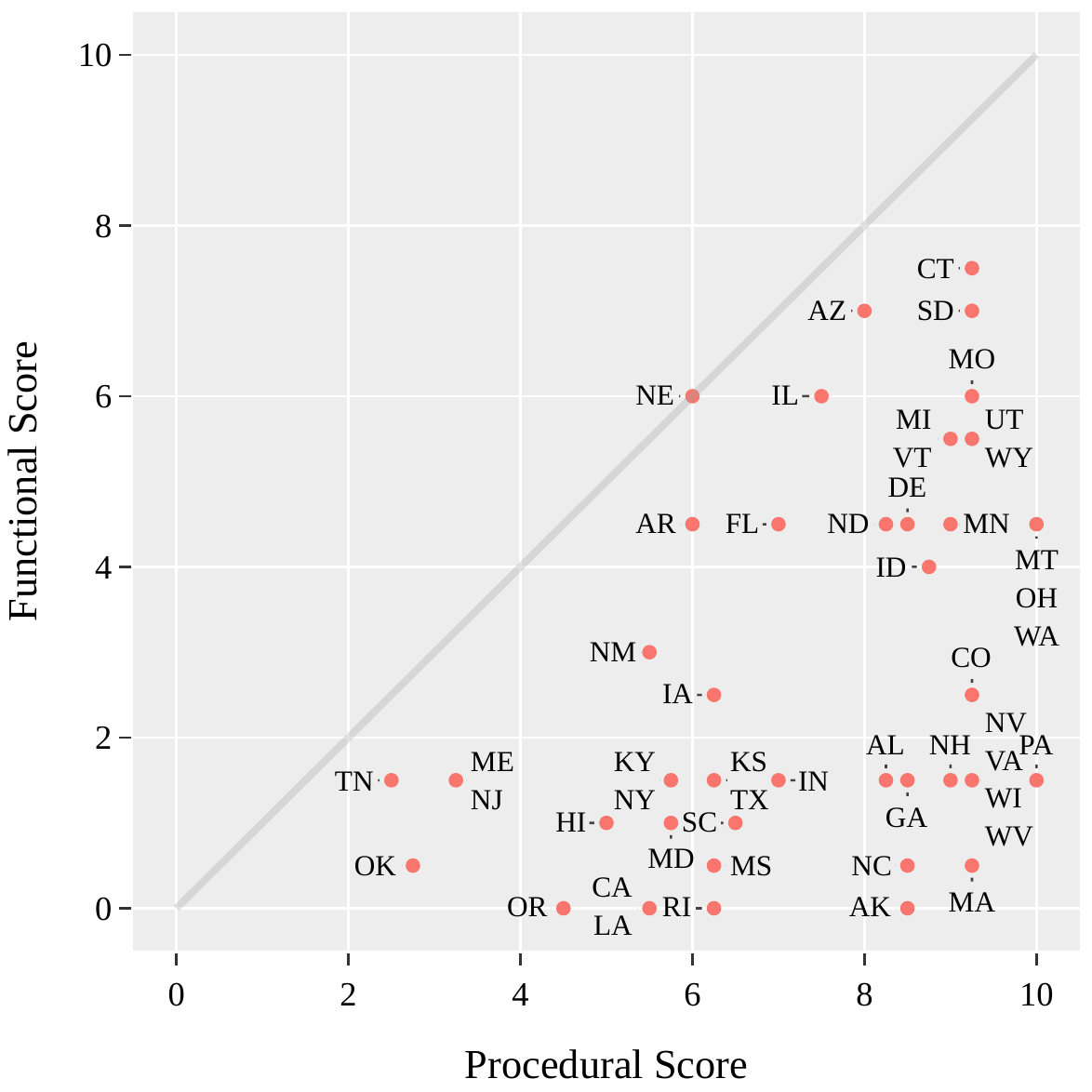}
    \caption{No state's functional score exceeds its procedural score, perhaps due to more limited data about functional aspects of testing. At most levels of procedural scores, states' functional scores spanned a wide range, with no strong correlation.}
    \label{fig:cross}
\end{figure}

\subsection{Performance by State}

When comparing states' overall L\&A testing practices, we find wide variation across both procedural and functional criteria. As illustrated in Figure~\ref{fig:maps}, this variation is not clearly explained by regionalism. However, the plot in Figure~\ref{fig:ordered} reveals several notable features in the distributions of states' scores.

Most obviously, procedural scores were much higher than functional scores. Again, this likely reflects both the relative scarcity of public documentation about functional aspects of L\&A testing and that our chosen functional criteria were somewhat aspirational. No states achieved perfect functional scores, but 4 states (Montana, Ohio, Pennsylvania, and Washington) achieved perfect procedural scores. Four other states could potentially achieve this benchmark but did not provide missing information we requested. Eleven more states clustered just shy of perfect procedural scores, of which 10 could achieve full points simply by making detailed L\&A procedures public (BP1)---potentially a zero-cost policy change.\looseness=-1
 
Notable relationships occur between certain criteria. For instance, concerning the scope of testing, only 2 states that are known to require testing every ballot (ST3) style do not also require testing every machine (ST1). It is much more common for states that require testing every machine to not require testing every ballot style (or remain silent), which 14 of 44 states did. This suggests that L\&A policymakers are more likely to be aware of the potential for problems that affect only specific machines than of issues that can affect only specific ballot styles.
 
The distribution of functional scores highlights further relationships. The largest cluster, at 1.5, are 16 states that require basic protections for voting targets (BP1) and overvotes (OP1) but meet no other functional criteria. Interestingly, many of these states employ similar statutory language, with little or no variation.\footnote{An additional state, South Carolina, adopted nearly the same statutory formula but with a small change that \emph{weakens} overvote protection, and so does not satisfy OP1.} Although we have so far been unable to identify the common source these states drew on, the situation suggests that providing stronger model legislation could be a fruitful way to encourage the adoption of improved L\&A practices.

At least 21 other states accomplish basic voting-target and overvote protections plus one or more of the stronger functional criteria. Most commonly, they require additional testing to detect transposed voting targets within contests (BP2), which 17 states do. Eight of these states accomplish no further functional criteria, resulting in a cluster at score 4.5. Five others also fully validate overvote thresholds (OP2), as do only 3 other states, indicating a strong correlation between these more rigorous testing policies. In a surprising contrast, although nondeterministic testing is comparably uncommon (only 8 states fully achieve either ND1 or ND2), practicing it does \emph{not} appear to be well correlated any of the other non-basic functional criteria. This may indicate that states have introduced nondeterministic testing haphazardly, rather than as the result of careful test-process design.

Considering both scoring categories together (Figure~\ref{fig:cross}), we see that although no state's functional score exceeds its procedural score, there is otherwise a wide range of functional scores at almost every level of procedural score. This may partly reflect limitations due to unresponsive states, but it may also suggest opportunities to better inform policymakers about ways to strengthen L\&A functionality, particularly in states in the lower-right corner of the figure, which have a demonstrated ability to develop procedurally robust testing requirements.

The overall national picture shows that every one of our evaluation criteria is satisfied by at least one state, indicating that even the most rigorous functional criteria are realistic to implement in practice. Several states have the potential to serve as models of best practice across most dimensions of L\&A testing, especially if procedural specifics are made readily accessible. In particular, Arizona and South Dakota each achieved full points in all but one criterion from each category, and Connecticut achieved the highest total score of any state under our metrics. We provide additional information and references regarding their procedures in our state-by-state summaries, found in Appendix~\ref{sec:states}.

\section{Discussion}
\label{sec:discussion}

Our findings support the need for strengthened L\&A procedures nationwide. Current practice has room for substantial improvement in both transparency and substance, and state policy should seek to realize this potential.

Election security researchers and practitioners should work together to establish normative standards for L\&A testing procedures and to draft model legislation to realize them. The precise mechanism for establishing this standard is beyond the scope of this paper, but a potential route would be for the National Institute of Standards and Technology (NIST) to issue L\&A testing guidelines. Under the Help America Vote Act (HAVA), NIST is charged with the design of voting system standards, in coordination with the U.S. Election Assistance Commission (EAC)~\cite{hava-2002}, and it has previously issued guidance for other aspects of election technology administration, such as cybersecurity and accessibility.

One challenge in the adoption of any technical standard is leaving safe and flexible opportunities for upward departure. It would be dangerous to lock in procedures that are later found to be insufficient, especially if every state would have to update its laws in response. For this reason, it is important that any L\&A policy changes allow some degree of flexibility involved for local jurisdictions. Too much flexibility, however, can weaken security guarantees even with the best of intentions. One clerk we spoke with in the preparation of this paper offhandedly told to us that she did not always follow the state requirement that every candidate in a contest receive a different number of votes, since in real elections ties could occur and she felt it was important to test that behavior too. Despite the well-meaning nature of this deviation, it \emph{decreased} the guarantees provided by her L\&A testing, since it meant the two candidates who tied in the test deck could have had their votes unnoticeably swapped. Clerks do not have the resources to rigorously analyze all ramifications of deviations from procedure, so latitude to deviate should be provided only where it cannot reduce the integrity of the process, such as in optional, additional phases of testing.

We leave to future work determining what model L\&A policies should look like. While the elements of transparency, openness, and security we considered in this paper are potential low-hanging fruit, there are other elements of successful L\&A practice that we did not measure or describe. For instance, testing policies should consider not only ballot scanners but also ballot marking devices (BMDs), which are computer kiosks that some voters use to mark and print their ballots. Most jurisdictions use BMDs primarily for voters with assistive needs, but some states require all in-person voters to use them~\cite{verifier}. Errors in BMD election definitions can lead to inaccurate results~\cite{dekalb-errors}, but carefully designed L\&A testing might reduce the incidence of such problems. Another example of an intervention that would have detected real-world issues in the past is ``end-to-end'' L\&A testing, where tabulator memory cards are loaded into the central election management system (EMS) and its result reports are checked against the test decks. One of the problems in Antrim County that caused it to report initially incorrect results in 2020 was an inconsistency between some tabulators and the EMS software, and ``end-to-end'' L\&A testing could have headed off this issue~\cite{antrim-report}.  

We do, however, recommend that future L\&A guidelines incorporate elements of nondeterministic testing. While our data shows that this practice is still quite rare in the U.S., using test decks that are unpredictable would make it more difficult to construct malicious election definitions that pass the testing procedure.

Election technology has evolved over time, but some L\&A testing practices still carry baggage from the past. For instance, functional requirements in many U.S. states are suited for detecting common problems with mechanical lever voting machines but less adept at uncovering common failure modes in modern computerized optical scanners, such as transposed voting targets.  Other nations, which may at this time be adopting optical scan equipment of their own, can learn from these standards and improve on them as they choose their own practices for the future. By applying careful scrutiny of existing process and incorporating the elements that most make sense in their own context, these polities can ensure their testing procedures are constructed in a way to meet their needs.

\section{Conclusion}

We performed the first detailed comparative analysis of L\&A testing procedures, based on a review of L\&A requirements and processes across all fifty U.S. states. Although L\&A testing can be a valuable tool for spotting common configuration errors and even certain kinds of low-tech attacks, our analysis shows that there is wide variation in how well states' testing requirements fulfill these prospects. We hope that our work can help rectify this by highlighting best practices as well as opportunities for improvement. Rigorous, transparent L\&A testing could also be a valuable tool for increasing public trust in elections, by giving voters a stronger basis for confidence that their votes will be counted correctly.

\section*{Acknowledgements}

We are grateful to the many election workers who corresponded with us to provide data for this study. We also thank our shepherd Nicole Goodman as well as Doug Jones, Dhanya Narayanan, Mike Specter, Drew Springall, and the anonymous reviewers. This work was supported by the Andrew Carnegie Fellowship, the U.S. National Science Foundation under grant CNS-1518888, and a gift from Microsoft.\looseness=-1

\setcounter{biburllcpenalty}{9000}
\printbibliography

\appendix
\section{State-by-State Practices}
\label{sec:states}

\newcommand{\AbbrevAlabama}{AL}\newcommand{\NameAlabama}{Alabama}\newcommand{\ProcScoreAlabama}{8.25}\newcommand{\ProcRankAlabama}{25}\newcommand{\ProcOrdAlabama}{th}\newcommand{\FuncScoreAlabama}{1.5}\newcommand{\FuncRankAlabama}{23}\newcommand{\FuncOrdAlabama}{rd}\newcommand{\TotScoreAlabama}{9.75}\newcommand{\TotRankAlabama}{28}\newcommand{\TotOrdAlabama}{th}\newcommand{\AbbrevAlaska}{AK}\newcommand{\NameAlaska}{Alaska}\newcommand{\ProcScoreAlaska}{8.5}\newcommand{\ProcRankAlaska}{21}\newcommand{\ProcOrdAlaska}{st}\newcommand{\FuncScoreAlaska}{0.0}\newcommand{\FuncRankAlaska}{46}\newcommand{\FuncOrdAlaska}{th}\newcommand{\TotScoreAlaska}{8.5}\newcommand{\TotRankAlaska}{32}\newcommand{\TotOrdAlaska}{nd}\newcommand{\AbbrevArizona}{AZ}\newcommand{\NameArizona}{Arizona}\newcommand{\ProcScoreArizona}{8.0}\newcommand{\ProcRankArizona}{27}\newcommand{\ProcOrdArizona}{th}\newcommand{\FuncScoreArizona}{7.0}\newcommand{\FuncRankArizona}{2}\newcommand{\FuncOrdArizona}{nd}\newcommand{\TotScoreArizona}{15.0}\newcommand{\TotRankArizona}{4}\newcommand{\TotOrdArizona}{th}\newcommand{\AbbrevArkansas}{AR}\newcommand{\NameArkansas}{Arkansas}\newcommand{\ProcScoreArkansas}{6.0}\newcommand{\ProcRankArkansas}{37}\newcommand{\ProcOrdArkansas}{th}\newcommand{\FuncScoreArkansas}{4.5}\newcommand{\FuncRankArkansas}{11}\newcommand{\FuncOrdArkansas}{th}\newcommand{\TotScoreArkansas}{10.5}\newcommand{\TotRankArkansas}{25}\newcommand{\TotOrdArkansas}{th}\newcommand{\AbbrevCalifornia}{CA}\newcommand{\NameCalifornia}{California}\newcommand{\ProcScoreCalifornia}{5.5}\newcommand{\ProcRankCalifornia}{42}\newcommand{\ProcOrdCalifornia}{nd}\newcommand{\FuncScoreCalifornia}{0.0}\newcommand{\FuncRankCalifornia}{46}\newcommand{\FuncOrdCalifornia}{th}\newcommand{\TotScoreCalifornia}{5.5}\newcommand{\TotRankCalifornia}{44}\newcommand{\TotOrdCalifornia}{th}\newcommand{\AbbrevColorado}{CO}\newcommand{\NameColorado}{Colorado}\newcommand{\ProcScoreColorado}{9.25}\newcommand{\ProcRankColorado}{5}\newcommand{\ProcOrdColorado}{th}\newcommand{\FuncScoreColorado}{2.5}\newcommand{\FuncRankColorado}{21}\newcommand{\FuncOrdColorado}{st}\newcommand{\TotScoreColorado}{11.75}\newcommand{\TotRankColorado}{18}\newcommand{\TotOrdColorado}{th}\newcommand{\AbbrevConnecticut}{CT}\newcommand{\NameConnecticut}{Connecticut}\newcommand{\ProcScoreConnecticut}{9.25}\newcommand{\ProcRankConnecticut}{5}\newcommand{\ProcOrdConnecticut}{th}\newcommand{\FuncScoreConnecticut}{7.5}\newcommand{\FuncRankConnecticut}{1}\newcommand{\FuncOrdConnecticut}{st}\newcommand{\TotScoreConnecticut}{16.75}\newcommand{\TotRankConnecticut}{1}\newcommand{\TotOrdConnecticut}{st}\newcommand{\AbbrevDelaware}{DE}\newcommand{\NameDelaware}{Delaware}\newcommand{\ProcScoreDelaware}{8.5}\newcommand{\ProcRankDelaware}{21}\newcommand{\ProcOrdDelaware}{st}\newcommand{\FuncScoreDelaware}{4.5}\newcommand{\FuncRankDelaware}{11}\newcommand{\FuncOrdDelaware}{th}\newcommand{\TotScoreDelaware}{13.0}\newcommand{\TotRankDelaware}{14}\newcommand{\TotOrdDelaware}{th}\newcommand{\AbbrevFlorida}{FL}\newcommand{\NameFlorida}{Florida}\newcommand{\ProcScoreFlorida}{7.0}\newcommand{\ProcRankFlorida}{29}\newcommand{\ProcOrdFlorida}{th}\newcommand{\FuncScoreFlorida}{4.5}\newcommand{\FuncRankFlorida}{11}\newcommand{\FuncOrdFlorida}{th}\newcommand{\TotScoreFlorida}{11.5}\newcommand{\TotRankFlorida}{19}\newcommand{\TotOrdFlorida}{th}\newcommand{\AbbrevGeorgia}{GA}\newcommand{\NameGeorgia}{Georgia}\newcommand{\ProcScoreGeorgia}{8.5}\newcommand{\ProcRankGeorgia}{21}\newcommand{\ProcOrdGeorgia}{st}\newcommand{\FuncScoreGeorgia}{1.5}\newcommand{\FuncRankGeorgia}{23}\newcommand{\FuncOrdGeorgia}{rd}\newcommand{\TotScoreGeorgia}{10.0}\newcommand{\TotRankGeorgia}{27}\newcommand{\TotOrdGeorgia}{th}\newcommand{\AbbrevHawaii}{HI}\newcommand{\NameHawaii}{Hawaii}\newcommand{\ProcScoreHawaii}{5.0}\newcommand{\ProcRankHawaii}{45}\newcommand{\ProcOrdHawaii}{th}\newcommand{\FuncScoreHawaii}{1.0}\newcommand{\FuncRankHawaii}{39}\newcommand{\FuncOrdHawaii}{th}\newcommand{\TotScoreHawaii}{6.0}\newcommand{\TotRankHawaii}{43}\newcommand{\TotOrdHawaii}{rd}\newcommand{\AbbrevIdaho}{ID}\newcommand{\NameIdaho}{Idaho}\newcommand{\ProcScoreIdaho}{8.75}\newcommand{\ProcRankIdaho}{20}\newcommand{\ProcOrdIdaho}{th}\newcommand{\FuncScoreIdaho}{4.0}\newcommand{\FuncRankIdaho}{19}\newcommand{\FuncOrdIdaho}{th}\newcommand{\TotScoreIdaho}{12.75}\newcommand{\TotRankIdaho}{15}\newcommand{\TotOrdIdaho}{th}\newcommand{\AbbrevIllinois}{IL}\newcommand{\NameIllinois}{Illinois}\newcommand{\ProcScoreIllinois}{7.5}\newcommand{\ProcRankIllinois}{28}\newcommand{\ProcOrdIllinois}{th}\newcommand{\FuncScoreIllinois}{6.0}\newcommand{\FuncRankIllinois}{4}\newcommand{\FuncOrdIllinois}{th}\newcommand{\TotScoreIllinois}{13.5}\newcommand{\TotRankIllinois}{12}\newcommand{\TotOrdIllinois}{th}\newcommand{\AbbrevIndiana}{IN}\newcommand{\NameIndiana}{Indiana}\newcommand{\ProcScoreIndiana}{7.0}\newcommand{\ProcRankIndiana}{29}\newcommand{\ProcOrdIndiana}{th}\newcommand{\FuncScoreIndiana}{1.5}\newcommand{\FuncRankIndiana}{23}\newcommand{\FuncOrdIndiana}{rd}\newcommand{\TotScoreIndiana}{8.5}\newcommand{\TotRankIndiana}{32}\newcommand{\TotOrdIndiana}{nd}\newcommand{\AbbrevIowa}{IA}\newcommand{\NameIowa}{Iowa}\newcommand{\ProcScoreIowa}{6.25}\newcommand{\ProcRankIowa}{32}\newcommand{\ProcOrdIowa}{nd}\newcommand{\FuncScoreIowa}{2.5}\newcommand{\FuncRankIowa}{21}\newcommand{\FuncOrdIowa}{st}\newcommand{\TotScoreIowa}{8.75}\newcommand{\TotRankIowa}{31}\newcommand{\TotOrdIowa}{st}\newcommand{\AbbrevKansas}{KS}\newcommand{\NameKansas}{Kansas}\newcommand{\ProcScoreKansas}{6.25}\newcommand{\ProcRankKansas}{32}\newcommand{\ProcOrdKansas}{nd}\newcommand{\FuncScoreKansas}{1.5}\newcommand{\FuncRankKansas}{23}\newcommand{\FuncOrdKansas}{rd}\newcommand{\TotScoreKansas}{7.75}\newcommand{\TotRankKansas}{35}\newcommand{\TotOrdKansas}{th}\newcommand{\AbbrevKentucky}{KY}\newcommand{\NameKentucky}{Kentucky}\newcommand{\ProcScoreKentucky}{5.75}\newcommand{\ProcRankKentucky}{39}\newcommand{\ProcOrdKentucky}{th}\newcommand{\FuncScoreKentucky}{1.5}\newcommand{\FuncRankKentucky}{23}\newcommand{\FuncOrdKentucky}{rd}\newcommand{\TotScoreKentucky}{7.25}\newcommand{\TotRankKentucky}{38}\newcommand{\TotOrdKentucky}{th}\newcommand{\AbbrevLouisiana}{LA}\newcommand{\NameLouisiana}{Louisiana}\newcommand{\ProcScoreLouisiana}{5.5}\newcommand{\ProcRankLouisiana}{42}\newcommand{\ProcOrdLouisiana}{nd}\newcommand{\FuncScoreLouisiana}{0.0}\newcommand{\FuncRankLouisiana}{46}\newcommand{\FuncOrdLouisiana}{th}\newcommand{\TotScoreLouisiana}{5.5}\newcommand{\TotRankLouisiana}{44}\newcommand{\TotOrdLouisiana}{th}\newcommand{\AbbrevMaine}{ME}\newcommand{\NameMaine}{Maine}\newcommand{\ProcScoreMaine}{3.25}\newcommand{\ProcRankMaine}{47}\newcommand{\ProcOrdMaine}{th}\newcommand{\FuncScoreMaine}{1.5}\newcommand{\FuncRankMaine}{23}\newcommand{\FuncOrdMaine}{rd}\newcommand{\TotScoreMaine}{4.75}\newcommand{\TotRankMaine}{46}\newcommand{\TotOrdMaine}{th}\newcommand{\AbbrevMaryland}{MD}\newcommand{\NameMaryland}{Maryland}\newcommand{\ProcScoreMaryland}{5.75}\newcommand{\ProcRankMaryland}{39}\newcommand{\ProcOrdMaryland}{th}\newcommand{\FuncScoreMaryland}{1.0}\newcommand{\FuncRankMaryland}{39}\newcommand{\FuncOrdMaryland}{th}\newcommand{\TotScoreMaryland}{6.75}\newcommand{\TotRankMaryland}{40}\newcommand{\TotOrdMaryland}{th}\newcommand{\AbbrevMassachusetts}{MA}\newcommand{\NameMassachusetts}{Massachusetts}\newcommand{\ProcScoreMassachusetts}{9.25}\newcommand{\ProcRankMassachusetts}{5}\newcommand{\ProcOrdMassachusetts}{th}\newcommand{\FuncScoreMassachusetts}{0.5}\newcommand{\FuncRankMassachusetts}{42}\newcommand{\FuncOrdMassachusetts}{nd}\newcommand{\TotScoreMassachusetts}{9.75}\newcommand{\TotRankMassachusetts}{28}\newcommand{\TotOrdMassachusetts}{th}\newcommand{\AbbrevMichigan}{MI}\newcommand{\NameMichigan}{Michigan}\newcommand{\ProcScoreMichigan}{9.0}\newcommand{\ProcRankMichigan}{16}\newcommand{\ProcOrdMichigan}{th}\newcommand{\FuncScoreMichigan}{5.5}\newcommand{\FuncRankMichigan}{7}\newcommand{\FuncOrdMichigan}{th}\newcommand{\TotScoreMichigan}{14.5}\newcommand{\TotRankMichigan}{7}\newcommand{\TotOrdMichigan}{th}\newcommand{\AbbrevMinnesota}{MN}\newcommand{\NameMinnesota}{Minnesota}\newcommand{\ProcScoreMinnesota}{9.0}\newcommand{\ProcRankMinnesota}{16}\newcommand{\ProcOrdMinnesota}{th}\newcommand{\FuncScoreMinnesota}{4.5}\newcommand{\FuncRankMinnesota}{11}\newcommand{\FuncOrdMinnesota}{th}\newcommand{\TotScoreMinnesota}{13.5}\newcommand{\TotRankMinnesota}{12}\newcommand{\TotOrdMinnesota}{th}\newcommand{\AbbrevMississippi}{MS}\newcommand{\NameMississippi}{Mississippi}\newcommand{\ProcScoreMississippi}{6.25}\newcommand{\ProcRankMississippi}{32}\newcommand{\ProcOrdMississippi}{nd}\newcommand{\FuncScoreMississippi}{0.5}\newcommand{\FuncRankMississippi}{42}\newcommand{\FuncOrdMississippi}{nd}\newcommand{\TotScoreMississippi}{6.75}\newcommand{\TotRankMississippi}{40}\newcommand{\TotOrdMississippi}{th}\newcommand{\AbbrevMissouri}{MO}\newcommand{\NameMissouri}{Missouri}\newcommand{\ProcScoreMissouri}{9.25}\newcommand{\ProcRankMissouri}{5}\newcommand{\ProcOrdMissouri}{th}\newcommand{\FuncScoreMissouri}{6.0}\newcommand{\FuncRankMissouri}{4}\newcommand{\FuncOrdMissouri}{th}\newcommand{\TotScoreMissouri}{15.25}\newcommand{\TotRankMissouri}{3}\newcommand{\TotOrdMissouri}{rd}\newcommand{\AbbrevMontana}{MT}\newcommand{\NameMontana}{Montana}\newcommand{\ProcScoreMontana}{10.0}\newcommand{\ProcRankMontana}{1}\newcommand{\ProcOrdMontana}{st}\newcommand{\FuncScoreMontana}{4.5}\newcommand{\FuncRankMontana}{11}\newcommand{\FuncOrdMontana}{th}\newcommand{\TotScoreMontana}{14.5}\newcommand{\TotRankMontana}{7}\newcommand{\TotOrdMontana}{th}\newcommand{\AbbrevNebraska}{NE}\newcommand{\NameNebraska}{Nebraska}\newcommand{\ProcScoreNebraska}{6.0}\newcommand{\ProcRankNebraska}{37}\newcommand{\ProcOrdNebraska}{th}\newcommand{\FuncScoreNebraska}{6.0}\newcommand{\FuncRankNebraska}{4}\newcommand{\FuncOrdNebraska}{th}\newcommand{\TotScoreNebraska}{12.0}\newcommand{\TotRankNebraska}{17}\newcommand{\TotOrdNebraska}{th}\newcommand{\AbbrevNevada}{NV}\newcommand{\NameNevada}{Nevada}\newcommand{\ProcScoreNevada}{9.25}\newcommand{\ProcRankNevada}{5}\newcommand{\ProcOrdNevada}{th}\newcommand{\FuncScoreNevada}{1.5}\newcommand{\FuncRankNevada}{23}\newcommand{\FuncOrdNevada}{rd}\newcommand{\TotScoreNevada}{10.75}\newcommand{\TotRankNevada}{21}\newcommand{\TotOrdNevada}{st}\newcommand{\AbbrevNewHampshire}{NH}\newcommand{\NameNewHampshire}{New Hampshire}\newcommand{\ProcScoreNewHampshire}{9.0}\newcommand{\ProcRankNewHampshire}{16}\newcommand{\ProcOrdNewHampshire}{th}\newcommand{\FuncScoreNewHampshire}{1.5}\newcommand{\FuncRankNewHampshire}{23}\newcommand{\FuncOrdNewHampshire}{rd}\newcommand{\TotScoreNewHampshire}{10.5}\newcommand{\TotRankNewHampshire}{25}\newcommand{\TotOrdNewHampshire}{th}\newcommand{\AbbrevNewJersey}{NJ}\newcommand{\NameNewJersey}{New Jersey}\newcommand{\ProcScoreNewJersey}{3.25}\newcommand{\ProcRankNewJersey}{47}\newcommand{\ProcOrdNewJersey}{th}\newcommand{\FuncScoreNewJersey}{1.5}\newcommand{\FuncRankNewJersey}{23}\newcommand{\FuncOrdNewJersey}{rd}\newcommand{\TotScoreNewJersey}{4.75}\newcommand{\TotRankNewJersey}{46}\newcommand{\TotOrdNewJersey}{th}\newcommand{\AbbrevNewMexico}{NM}\newcommand{\NameNewMexico}{New Mexico}\newcommand{\ProcScoreNewMexico}{5.5}\newcommand{\ProcRankNewMexico}{42}\newcommand{\ProcOrdNewMexico}{nd}\newcommand{\FuncScoreNewMexico}{3.0}\newcommand{\FuncRankNewMexico}{20}\newcommand{\FuncOrdNewMexico}{th}\newcommand{\TotScoreNewMexico}{8.5}\newcommand{\TotRankNewMexico}{32}\newcommand{\TotOrdNewMexico}{nd}\newcommand{\AbbrevNewYork}{NY}\newcommand{\NameNewYork}{New York}\newcommand{\ProcScoreNewYork}{5.75}\newcommand{\ProcRankNewYork}{39}\newcommand{\ProcOrdNewYork}{th}\newcommand{\FuncScoreNewYork}{1.5}\newcommand{\FuncRankNewYork}{23}\newcommand{\FuncOrdNewYork}{rd}\newcommand{\TotScoreNewYork}{7.25}\newcommand{\TotRankNewYork}{38}\newcommand{\TotOrdNewYork}{th}\newcommand{\AbbrevNorthCarolina}{NC}\newcommand{\NameNorthCarolina}{North Carolina}\newcommand{\ProcScoreNorthCarolina}{8.5}\newcommand{\ProcRankNorthCarolina}{21}\newcommand{\ProcOrdNorthCarolina}{st}\newcommand{\FuncScoreNorthCarolina}{0.5}\newcommand{\FuncRankNorthCarolina}{42}\newcommand{\FuncOrdNorthCarolina}{nd}\newcommand{\TotScoreNorthCarolina}{9.0}\newcommand{\TotRankNorthCarolina}{30}\newcommand{\TotOrdNorthCarolina}{th}\newcommand{\AbbrevNorthDakota}{ND}\newcommand{\NameNorthDakota}{North Dakota}\newcommand{\ProcScoreNorthDakota}{8.25}\newcommand{\ProcRankNorthDakota}{25}\newcommand{\ProcOrdNorthDakota}{th}\newcommand{\FuncScoreNorthDakota}{4.5}\newcommand{\FuncRankNorthDakota}{11}\newcommand{\FuncOrdNorthDakota}{th}\newcommand{\TotScoreNorthDakota}{12.75}\newcommand{\TotRankNorthDakota}{15}\newcommand{\TotOrdNorthDakota}{th}\newcommand{\AbbrevOhio}{OH}\newcommand{\NameOhio}{Ohio}\newcommand{\ProcScoreOhio}{10.0}\newcommand{\ProcRankOhio}{1}\newcommand{\ProcOrdOhio}{st}\newcommand{\FuncScoreOhio}{4.5}\newcommand{\FuncRankOhio}{11}\newcommand{\FuncOrdOhio}{th}\newcommand{\TotScoreOhio}{14.5}\newcommand{\TotRankOhio}{7}\newcommand{\TotOrdOhio}{th}\newcommand{\AbbrevOklahoma}{OK}\newcommand{\NameOklahoma}{Oklahoma}\newcommand{\ProcScoreOklahoma}{2.75}\newcommand{\ProcRankOklahoma}{49}\newcommand{\ProcOrdOklahoma}{th}\newcommand{\FuncScoreOklahoma}{0.5}\newcommand{\FuncRankOklahoma}{42}\newcommand{\FuncOrdOklahoma}{nd}\newcommand{\TotScoreOklahoma}{3.25}\newcommand{\TotRankOklahoma}{50}\newcommand{\TotOrdOklahoma}{th}\newcommand{\AbbrevOregon}{OR}\newcommand{\NameOregon}{Oregon}\newcommand{\ProcScoreOregon}{4.5}\newcommand{\ProcRankOregon}{46}\newcommand{\ProcOrdOregon}{th}\newcommand{\FuncScoreOregon}{0.0}\newcommand{\FuncRankOregon}{46}\newcommand{\FuncOrdOregon}{th}\newcommand{\TotScoreOregon}{4.5}\newcommand{\TotRankOregon}{48}\newcommand{\TotOrdOregon}{th}\newcommand{\AbbrevPennsylvania}{PA}\newcommand{\NamePennsylvania}{Pennsylvania}\newcommand{\ProcScorePennsylvania}{10.0}\newcommand{\ProcRankPennsylvania}{1}\newcommand{\ProcOrdPennsylvania}{st}\newcommand{\FuncScorePennsylvania}{1.5}\newcommand{\FuncRankPennsylvania}{23}\newcommand{\FuncOrdPennsylvania}{rd}\newcommand{\TotScorePennsylvania}{11.5}\newcommand{\TotRankPennsylvania}{19}\newcommand{\TotOrdPennsylvania}{th}\newcommand{\AbbrevRhodeIsland}{RI}\newcommand{\NameRhodeIsland}{Rhode Island}\newcommand{\ProcScoreRhodeIsland}{6.25}\newcommand{\ProcRankRhodeIsland}{32}\newcommand{\ProcOrdRhodeIsland}{nd}\newcommand{\FuncScoreRhodeIsland}{0.0}\newcommand{\FuncRankRhodeIsland}{46}\newcommand{\FuncOrdRhodeIsland}{th}\newcommand{\TotScoreRhodeIsland}{6.25}\newcommand{\TotRankRhodeIsland}{42}\newcommand{\TotOrdRhodeIsland}{nd}\newcommand{\AbbrevSouthCarolina}{SC}\newcommand{\NameSouthCarolina}{South Carolina}\newcommand{\ProcScoreSouthCarolina}{6.5}\newcommand{\ProcRankSouthCarolina}{31}\newcommand{\ProcOrdSouthCarolina}{st}\newcommand{\FuncScoreSouthCarolina}{1.0}\newcommand{\FuncRankSouthCarolina}{39}\newcommand{\FuncOrdSouthCarolina}{th}\newcommand{\TotScoreSouthCarolina}{7.5}\newcommand{\TotRankSouthCarolina}{37}\newcommand{\TotOrdSouthCarolina}{th}\newcommand{\AbbrevSouthDakota}{SD}\newcommand{\NameSouthDakota}{South Dakota}\newcommand{\ProcScoreSouthDakota}{9.25}\newcommand{\ProcRankSouthDakota}{5}\newcommand{\ProcOrdSouthDakota}{th}\newcommand{\FuncScoreSouthDakota}{7.0}\newcommand{\FuncRankSouthDakota}{2}\newcommand{\FuncOrdSouthDakota}{nd}\newcommand{\TotScoreSouthDakota}{16.25}\newcommand{\TotRankSouthDakota}{2}\newcommand{\TotOrdSouthDakota}{nd}\newcommand{\AbbrevTennessee}{TN}\newcommand{\NameTennessee}{Tennessee}\newcommand{\ProcScoreTennessee}{2.5}\newcommand{\ProcRankTennessee}{50}\newcommand{\ProcOrdTennessee}{th}\newcommand{\FuncScoreTennessee}{1.5}\newcommand{\FuncRankTennessee}{23}\newcommand{\FuncOrdTennessee}{rd}\newcommand{\TotScoreTennessee}{4.0}\newcommand{\TotRankTennessee}{49}\newcommand{\TotOrdTennessee}{th}\newcommand{\AbbrevTexas}{TX}\newcommand{\NameTexas}{Texas}\newcommand{\ProcScoreTexas}{6.25}\newcommand{\ProcRankTexas}{32}\newcommand{\ProcOrdTexas}{nd}\newcommand{\FuncScoreTexas}{1.5}\newcommand{\FuncRankTexas}{23}\newcommand{\FuncOrdTexas}{rd}\newcommand{\TotScoreTexas}{7.75}\newcommand{\TotRankTexas}{35}\newcommand{\TotOrdTexas}{th}\newcommand{\AbbrevUtah}{UT}\newcommand{\NameUtah}{Utah}\newcommand{\ProcScoreUtah}{9.25}\newcommand{\ProcRankUtah}{5}\newcommand{\ProcOrdUtah}{th}\newcommand{\FuncScoreUtah}{5.5}\newcommand{\FuncRankUtah}{7}\newcommand{\FuncOrdUtah}{th}\newcommand{\TotScoreUtah}{14.75}\newcommand{\TotRankUtah}{5}\newcommand{\TotOrdUtah}{th}\newcommand{\AbbrevVermont}{VT}\newcommand{\NameVermont}{Vermont}\newcommand{\ProcScoreVermont}{9.0}\newcommand{\ProcRankVermont}{16}\newcommand{\ProcOrdVermont}{th}\newcommand{\FuncScoreVermont}{5.5}\newcommand{\FuncRankVermont}{7}\newcommand{\FuncOrdVermont}{th}\newcommand{\TotScoreVermont}{14.5}\newcommand{\TotRankVermont}{7}\newcommand{\TotOrdVermont}{th}\newcommand{\AbbrevVirginia}{VA}\newcommand{\NameVirginia}{Virginia}\newcommand{\ProcScoreVirginia}{9.25}\newcommand{\ProcRankVirginia}{5}\newcommand{\ProcOrdVirginia}{th}\newcommand{\FuncScoreVirginia}{1.5}\newcommand{\FuncRankVirginia}{23}\newcommand{\FuncOrdVirginia}{rd}\newcommand{\TotScoreVirginia}{10.75}\newcommand{\TotRankVirginia}{21}\newcommand{\TotOrdVirginia}{st}\newcommand{\AbbrevWashington}{WA}\newcommand{\NameWashington}{Washington}\newcommand{\ProcScoreWashington}{10.0}\newcommand{\ProcRankWashington}{1}\newcommand{\ProcOrdWashington}{st}\newcommand{\FuncScoreWashington}{4.5}\newcommand{\FuncRankWashington}{11}\newcommand{\FuncOrdWashington}{th}\newcommand{\TotScoreWashington}{14.5}\newcommand{\TotRankWashington}{7}\newcommand{\TotOrdWashington}{th}\newcommand{\AbbrevWestVirginia}{WV}\newcommand{\NameWestVirginia}{West Virginia}\newcommand{\ProcScoreWestVirginia}{9.25}\newcommand{\ProcRankWestVirginia}{5}\newcommand{\ProcOrdWestVirginia}{th}\newcommand{\FuncScoreWestVirginia}{1.5}\newcommand{\FuncRankWestVirginia}{23}\newcommand{\FuncOrdWestVirginia}{rd}\newcommand{\TotScoreWestVirginia}{10.75}\newcommand{\TotRankWestVirginia}{21}\newcommand{\TotOrdWestVirginia}{st}\newcommand{\AbbrevWisconsin}{WI}\newcommand{\NameWisconsin}{Wisconsin}\newcommand{\ProcScoreWisconsin}{9.25}\newcommand{\ProcRankWisconsin}{5}\newcommand{\ProcOrdWisconsin}{th}\newcommand{\FuncScoreWisconsin}{1.5}\newcommand{\FuncRankWisconsin}{23}\newcommand{\FuncOrdWisconsin}{rd}\newcommand{\TotScoreWisconsin}{10.75}\newcommand{\TotRankWisconsin}{21}\newcommand{\TotOrdWisconsin}{st}\newcommand{\AbbrevWyoming}{WY}\newcommand{\NameWyoming}{Wyoming}\newcommand{\ProcScoreWyoming}{9.25}\newcommand{\ProcRankWyoming}{5}\newcommand{\ProcOrdWyoming}{th}\newcommand{\FuncScoreWyoming}{5.5}\newcommand{\FuncRankWyoming}{7}\newcommand{\FuncOrdWyoming}{th}\newcommand{\TotScoreWyoming}{14.75}\newcommand{\TotRankWyoming}{5}\newcommand{\TotOrdWyoming}{th}

\newenvironment{State}[1]{%
    \medskip\noindent\textbf{\expandafter\csname Name#1\endcsname\ (\expandafter\csname Abbrev#1\endcsname)}\hfill%
    \emph{Proc.}: \expandafter\csname ProcScore#1\endcsname\ (\csname ProcRank#1\endcsname$\textsuperscript{\expandafter\csname ProcOrd#1\endcsname}$)\enspace%
    \emph{Func.}: \csname FuncScore#1\endcsname\ (\csname FuncRank#1\endcsname$\textsuperscript{\expandafter\csname FuncOrd#1\endcsname}$)\enspace%
    \emph{Total}: \csname TotScore#1\endcsname\ (\csname TotRank#1\endcsname$\textsuperscript{\expandafter\csname TotOrd#1\endcsname}$)%
    \nopagebreak[4]\smallskip\noindent\\%
}

Here we summarize notable features of each state's L\&A testing practices with respect to our evaluation criteria. We list each state's score and rank under the procedural and functional criteria (each /10 points) and their total (/20 points).\medskip

\begin{State}{Alabama}
Alabama's L\&A~\cite{alabama-e-voting-admin-code} public testing requirements vary by class of device. Direct-recording electronic voting devices (DREs), which are no longer used~\cite{verifier}, had to be tested at an event open to the general public. Testing of the state's current optical scan machines occurs in two phases: all devices are tested in a process observable by candidates or their representatives, and a subset of the devices is tested again in view of the general public. Interestingly, the state requires each candidate receive a minimum of \emph{two} votes, although there is no requirement that candidates within a contest receive a \emph{different} number of votes. Alabama permits local jurisdictions to implement more stringent testing practices in addition to required testing, so practices could be independently improved by local jurisdictions.\looseness=-1
\end{State}

\begin{State}{Alaska}
Documents describing Alaska's L\&A procedures are not publicly accessible, but the state's Division of Elections provided them in heavily redacted form in response to our requests~\cite{alaska-lat-testing-early-vote, alaska-lat-testing-voting-tablet, alaska-lat-testing-vvpat, alaska-democracy-suite-election}. Although Alaska requires each voting machine to be publicly tested, the redactions precluded our finding evidence responsive to most of our other criteria, resulting in a low score. Alaska has not responded to a request for clarification. 
\end{State}

\begin{State}{Arizona}
Arizona uses a random number generator to mark test ballots, creating a nondeterministic test deck that would be difficult to predict and earning the state full credit for ND2. The state also firmly bounds the overvote thresholds by voting at least one ballot for the maximum number of permissible choices in each contest and then exactly one additional vote than what is allowable~\cite{arizona-elections-procedures-manual}. Lastly, the state permits observers from local political parties to contribute their own votes to a test ballot. The state could earn a perfect score overall by testing each ballot style and ensuring candidates within a contest receive distinct vote totals.
\end{State}

\begin{State}{Arkansas}
Arkansas allows the public to observe the testing of all devices prior to an election. The state ensures every candidate receives at least one vote and that at least one overvoted ballot must be cast~\cite{arkansas-procedures-manual}. Its procedures can be strengthened by specifying proper overvote validation practice, ensuring different vote totals for each candidate in a contest, and introducing nondeterministic test elements. Arkansas has not responded to requests for additional information.
\end{State}

\begin{State}{California}
California's Election Code § 15000~\cite{california-election-code} requires the state to conduct logic and accuracy testing. The state, however, relies on  vendor-provided procedures that vary from machine to machine instead of implementing statewide requirements that are independent of the vendor. We assessed that this reliance on vendor-provided testing material does not satisfy our functional criteria, since even if every vendor's manual happens to meet a particular requirement now, updated manuals could weaken these provisions without any conscious regulatory action. Accordingly, the state scores lower than it likely would have otherwise.
\end{State}

\begin{State}{Colorado}
Colorado has robust procedural L\&A policies that are publicly applied to every machine~\cite{colorado-crs}. Still, Colorado has room to improve in their functional protections. The state currently does not require vote totals to differ between various candidates for an office, but instead encourages it by providing a ``ballot position calculator." Furthermore, Colorado does not require a strict test of overvote protection. Most of the information for our assessment was provided via communication with the state as there is no statewide procedure document. 
\end{State}

\begin{State}{Connecticut}
Connecticut's public statues require L\&A testing~\cite{connecticut-general-statutes}. Communication with the state election division revealed that the state comes close to fulfilling almost all of our criteria and made it the highest-scoring state overall. The main opportunity we find for improvement would be to strengthen nondeterministic testing. Connecticut already earns partial credit for ND2 since its procedures demonstrate an understanding of the importance of randomized testing, even though the source of randomness is not specified. To earn a perfect functional score, the state should require the use of a random number generator to mark some number of test ballots and also allow public observers to mark and cast test ballots arbitrarily. 
\end{State}

\begin{State}{Delaware}
Delaware's L\&A policies are unclear based on its statutes~\cite{delaware-code}, but a FOIA request for testing policies submitted at our behest by a Delaware resident yielded more informative documents~\cite{ballot-procedures-foia}. We therefore consider criterion RT1 to be met, although the process of obtaining the documentation was needlessly difficult.  In addition to a lack of transparency around their L\&A documentation, it is unclear whether the state allows local jurisdictions to practice more stringent testing requirements. However, all other procedural testing criteria were met. In terms of functional protections, both basic protections were included as part of testing requirements. Overvote protections are not strictly checked; test decks include ballots where all choices are marked for each office but do not include ones marked with exactly one more vote than is permissible. Additionally, election officials are not required to mark test ballots randomly. 
\end{State}

\begin{State}{Florida}
Florida tests all tabulators during a “100\% Logic and Accuracy Testing” event and a sample during a public L\&A test~\cite{florida-guidelines}. The state meets all procedural testing criteria except for requiring all machines to be tested publicly and with each ballot style. Florida recognizes that some counties use a traditional 1-2-3 test deck pattern (satisfying BP2) but notes that it is not the most accurate way to verify that ballots are being counted correctly. To supplement this testing mechanism, the state encourages (although does not require) the creation of an ```enhanced test deck with non-traditional vote patterns."
\end{State}

\begin{State}{Georgia}
Georgia maintains a step-by-step document that describes how to implement its L\&A rules~\cite{georgia-procedures}. The procedures, however, do not meet many criteria including requiring testing of  all ballot styles and ensuring candidates for an office have different vote totals. Additionally, the procedures could not be located from the state's election website.
\end{State}

\begin{State}{Hawaii}
Hawaii’s L\&A procedures~\cite{hawaii-counting-center-manual} differ substantially from most other states. The state does not permit the general public to observe testing. Instead, ``Official Observers" must be designated by a political party, news media organization, or chief election officer to ``serve as the `eyes and ears' of the public." In practice, ``Official Observers" are quasi election officials who are tasked with conducting testing however they choose. Given Hawaii's unique situation, we assigned the state partial credit for ND1. However, since testing is not open to members of the general public unless they become ``Official Observers," Hawaii did not earn credit for RT3. Hawaii earned full credit for requiring all tabulators to be tested, but it failed to meet other key criteria including ensuring that each candidate receives at least one vote and that all ballot styles are included during testing.
\end{State}

\begin{State}{Idaho}
Although Idaho does not have a statewide document dedicated to explaining the state's testing procedures, communication with state officials revealed that Idaho meets all functional criteria except for requiring proper overvote validation (BP2) and for election officials to mark ballots using a source of randomness (ND2).
\end{State}

\begin{State}{Illinois}
The Illinois L\&A testing best practice guide~\cite{illinois-best-practices} fulfills many functional and procedural criteria. Importantly, the state tactfully  juxtaposes an example unsatisfactory test with an example satisfactory test in way that efficiently conveys its L\&A requirements---a method that states hoping to improve procedural clarity should consider. Illinois would benefit from introducing nondeterministic elements to its L\&A testing and requiring that all ballot styles be tested on each tabulator. 
\end{State}

\begin{State}{Indiana}
Indiana distinguishes between public testing and logic and accuracy testing in its procedure manual~\cite{indiana-election-admin-manual}, requiring the former and leaving the latter undefined. What the state terms ``public testing,'' however, is analogous to other states' L\&A. While the state requires testing of all optical scan (``ballot card'') tabulators, only 5\% of DRE voting systems must be tested. Since much of Indiana still uses DREs~\cite{verifier}, it should strongly consider requiring all devices to undergo testing prior to each election as well as introducing explicit requirements regarding basic protections, overvote threshold validation, and nondeterminism. Indiana does require that all ballot styles be tested on each device subject to testing. 
\end{State}

\begin{State}{Iowa}
 Iowa has limited public documentation of its election procedures and has not responded to any of our email inquiries regarding its testing practices. Therefore, the state was scored in accordance with what could be found in its Election Code \cite{iowa-code} and an election security informational video on its website~\cite{iowa-election-video}.
\end{State}

\begin{State}{Kansas}
Kansas includes L\&A testing requirements as part of its state code, \S 25-4411~\cite{kansas-statutes}. This statute requires public testing with a test deck that includes at least one overvoted ballot for each machine. Kansas also requires that L\&A testing be open to the public and that all machines are tested. This gives the state a somewhat better procedural testing score than it would have had otherwise, but there are still many improvements that can be made, such as specifying a requirement for the testing of each ballot style and detection of transposed targets. The state's election division has not responded to our inquiries regarding testing requirements and so has been scored purely on the basis of public information.
\end{State}

\begin{State}{Kentucky}
Kentucky does not have any statewide procedural L\&A document, but the state does have L\&A requirements under its administrative code~\cite{kentucky-code}. Kentucky earned points for requiring all of its voting equipment to undergo testing; however, the state does not require public L\&A testing. Instead, it permits a representative from each political party and representatives of news media to be present at a ``Public Examination'' as the county elections board ensures: ballots are properly arranged, counters are set to zero, equipment is locked, and that the equipment's assigned serial number is recorded. 
\end{State}

\begin{State}{Louisiana}
Louisiana Election Code~\cite{louisiana-election-code} requires testing of every machine before each election, but it does not appear to establish any minimum standards for what this testing entails. Unusually, only Louisiana citizens are permitted to observe testing, so the state earns only partial credit for RT3. The state still earned full credit for ST2 because it ensures that the permitted public observers may witness the preparation and testing of each machine.
\end{State}

\begin{State}{Maine}
Found in 1.21--A M.R.S.A~\cite{maine-election-code}, Maine’s L\&A policy has several opportunities for improvement. It is unclear whether each tabulator is required to be tested, whether all ballot styles are included in testing, whether two candidates from the same contest can receive the same number of votes, and so forth. Observers are also not permitted to test ballots and machines themselves. Maine should consider producing publicly accessible procedure or guideline documents that expand upon its policy. Maine's Bureau of Corporations, Elections and Commission has not responded to our email inquiries regarding its testing procedures. 
\end{State}

\begin{State}{Maryland}
Described in Maryland COMAR (33.10.01.14--16)~\cite{maryland-code}, the state’s L\&A testing consists of a ``Pre-Election Test" of all tabulators.  It is followed by a ``Public Demonstration" that is limited to attendance by one representative of each political party and independent candidate. The public demonstration only consists of documentation completed during the pre-test and an overview of the testing process. Maryland should consider permitting the general public to observe ``Pre-Election Testing" and creating a readily accessible statewide procedural document so that the public can make informed observations. 
\end{State}

\begin{State}{Massachusetts}
Massachusetts does not have an approved statewide document regarding L\&A procedures, but it does have regulations on the subject~\cite{massachusetts-regulations}. Communication with state officials uncovered that the state's internal guidelines meet most procedural testing criteria. However, the state performs poorly in terms of test functionality, where the only criterion it meets is basic overvote protection (OP1).
\end{State}

\begin{State}{Michigan}
Michigan’s L\&A procedures are documented in a public manual~\cite{michigan-test-procedure-manual} and consist of a preliminary accuracy test in which all tabulators and BMDs are tested, as well as a public test in which only a sample of tabulators are required to be tested. Each candidate in a contest must receive a different number of votes, and overvoted ballots must be cast. Additionally, observers can mark and cast test ballots. We recommend that Michigan further ensure that all machines undergo public testing, that overvote thresholds are fully validated, and that some ballots are marked using a source of randomness.  
\end{State}

\begin{State}{Minnesota}
The state's public L\&A guidelines~\cite{minnesota-voting-equipment-test-guide} identify two testing events---a preliminary test in which all ballot counters are required to be tested and a public accuracy test in which only a sample is tested. Minnesota includes step-by-step instructions for creating test deck spreadsheets and provides samples that jurisdiction can utilize. The state does well in ensuring that its guidelines protect against problems identified in the basic protections category. The state could benefit from introducing nondeterministic elements to its testing as well as requiring election jurisdictions to fully validate overvote thresholds. 
\end{State}

\begin{State}{Mississippi}
Mississippi’s L\&A requirements can be found in its state code~\cite{mississippi-election-handbook}, which only requires that each machine be publicly tested for basic overvote rejection (OP1). The state has not responded to an inquiry regarding other procedural and functional elements of its testing.
\end{State}

\begin{State}{Missouri}
Missouri's L\&A policy is defined in 15 CSR 30--10.040~\cite{missouri-regulations}. Testing ensures that each candidate for an office receives a distinct and nonzero number of votes. Missouri is one of the few states that require full validation of overvote thresholds, with a requirement as follows: ``In situations where a voter can legally vote for more than one person for an office, at least one card shall be voted for the maximum number of allowable candidates; one card shall [then] be marked to have one more vote for each candidate or question than is allowable." This language could serve as a model for other states. We recommend introducing nondeterministic testing but commend the state for its unusually clear functional policies.\looseness=-1
\end{State}

\begin{State}{Montana}
Under Montana’s detailed public L\&A procedures~\cite{montana-procedures-guide}, testing is divided into three dimensions---functional, diagnostic, and physical---and fulfills almost all of our criteria. Montana could strengthen its procedures by fully bounding overvote thresholds and introducing forms of nondeterministic testing.
\end{State}

\begin{State}{Nebraska}
Nebraska’s logic and accuracy test is commonly referred to as a ``Mock Election." All tabulators undergo three independent tests using different test decks---one by the election official, one by the chief election commissioner, and one by the person who installed the election definition on the voting device. The state’s Election Act~\cite{nebraska-statutes} does not indicate that testing observation is open to members of the public; however, its test procedures meet all other criteria in the scope of testing, basic protections, and overvote protection categories. Nebraska could further improve its test functionality by requiring nondeterministic testing.
\end{State}
\begin{State}{Nevada}
In Nevada, L\&A testing is primarily described by state law~\cite{nevada-statutes}. All tabulators must be tested before an election in a process that can be observed by the general public, and local jurisdictions are allowed to exceed state testing requirements. Nevada would benefit from creating a public, statewide guide that describes the steps necessary to properly conduct L\&A testing.
\end{State}

\begin{State}{NewHampshire}
New Hampshire's \emph{Election Procedure Manual}~\cite{new-hampshire-procedure} meets all scope of testing requirements but otherwise does not address several key criteria, including the more rigorous of basic and overvote protections. Although the state broadly requires that election officials mark ballots with ``as many combinations as possible,'' we deem this to fall short of satisfying our nondeterministic testing criteria.\footnote{Taken literally, this is intractable. There are $2^n$ ways to mark $n$ voting targets; an election with 50 candidates would require approximately as many ballots as there are grains of sand on earth.\looseness=-1} Calling for some test ballots to be marked truly at random would strengthen this provision.\looseness=-1
\end{State}

\begin{State}{NewJersey}
New Jersey's L\&A policy is described by state statutes 19:53A-8~\cite{new-jersey-statutes}. Testing is conducted publicly and includes one vote for each candidate as well as the casting of overvoted ballots. State law was unclear regarding other key aspects of testing, including whether all machines are to be tested prior to each election. The state did not respond to additional requests for information.
\end{State}

\begin{State}{NewMexico}
New Mexico has a handbook of relevant election code and legislation, but no procedure document~\cite{new-mexico-election-handbook}. While this legislation requires public testing before each election, no other information was present relevant to our criteria. Further communication with state officials revealed that only ``party and organization representatives, election observers and candidates'' are allowed to observe testing. We did not receive a response on whether all ballot styles are tested on each tabulator. Functionally, the state's testing implements both overvote protections but only the first basic protection (BP1). 
\end{State}

\begin{State}{NewYork}
New York has a relatively minimal set of requirements defining their L\&A procedure~\cite{new-york-election-law}, which they term ``Prequalification Testing.'' Under these rules, all tabulators and all ballot styles are tested, but other important properties are not met: testing is not open to the public, overvote thresholds are not well-bounded, and multiple candidates in the same contest may receive an equal number of votes. We recommend addressing these shortfalls and introducing nondeterministic testing elements. The New York State Board of Elections did not respond to an email inquiry requesting additional information. 
\end{State}

\begin{State}{NorthCarolina}
North Carolina's L\&A testing is briefly described on the state's election website~\cite{north-carolina-accurate-elections}; we were able to obtain the state's procedures via correspondence with the State Board of Elections. The step-by-step list meets most of our procedural criteria but almost none of our functional criteria.
\end{State}

\begin{State}{NorthDakota}
Email communication with state election officials provided a great deal of insight regarding North Dakota's L\&A testing~\cite{north-dakota-regulations}. The state met the more advanced criteria requiring that all ballot styles be tested on every machine (ST3) and that a different numbers of votes be assigned to each option for each contest. It does not, however, require validating overvote thresholds or any nondeterministic testing.\looseness=-1  
\end{State}

\begin{State}{Ohio}
Ohio's L\&A practices~\cite{ohio-election-manual} indicate that state election officials are required to test tabulating computer programs prior to a given election and tabulating equipment prior to its use to count ballots. In addition to fulfilling each scope of testing and basic protections criterion, the state met all of our procedural criteria. The state would earn a perfect combined scored if it were to specify a requirements that election officials fully validate overvote thresholds and introduce nondeterministic elements into the test process. 
\end{State}

\begin{State}{Oklahoma}
State law in Oklahoma contains some provisions related to testing but leaves significant gaps relative to our criteria. When we contacted the State Election Board to request more information, they provided a page from the state's \textit{Uniform Election Reference}~\cite{oklahoma-election-reference}. This still failed to answer many of our questions, but contained some language which suggests local election officials may have access to additional private documents further defining L\&A requirements. The state did not answer follow-up requests for additional information, so we were only able to use public information and the short excerpt of the testing documentation when scoring the state's practice. 
\end{State}

\begin{State}{Oregon}
Oregon conducts all elections by mail and so defines L\&A policy in its \textit{Vote by Mail Procedure Manual}~\cite{oregon-vbm-manual}. The state’s L\&A testing is divided into a preparatory test in which all tabulators are required to be tested and a public certification test in which only a sample is tested. Observation of public testing is limited to one representative of each party and each nonpartisan candidate or their designated representative. There is no indication that any of the criteria in basic protections, overvote protection, and nondeterminism are met.
\end{State}

\begin{State}{Pennsylvania}
Pennsylvania has robust testing requirements~\cite{pennsylvania-directive} with an appropriate scope and excellent transparency. The state also has several good functional recommendations, including encouraging jurisdictions to assign a different number of votes to each candidate in a contest. We recommend that Pennsylvania turn its recommended practices into requirements. The state should also introduce elements of nondeterminism into its testing practices.
\end{State}

\begin{State}{RhodeIsland}
Rhode Island’s State Board of Elections works in conjunction with the voting equipment vendor to publicly test all tabulators before each election. Its L\&A policy~\cite{rhode-island-general-laws}, however, is very vague regarding our functional criteria, which led to the state earning no credit for that category. The Board did not respond to our correspondence seeking additional information.
\end{State}

\begin{State}{SouthCarolina}
Even though South Carolina's State Election Commission provided us with an excerpt of the state's L\&A procedures~\cite{south-carolina-accumulation-guide}, the document was so heavily redacted that we were not able to obtain any useful information relative to our criteria. Instead, we relied on state code \cite{south-carolina-code}, which uses similar language to several other states but is unique in that it can be satisfied by the inclusion of overvoted \textit{or} undervoted ballots and thus fails to meet OP1. The State Election Commission did not responded to our request for additional information. 
\end{State}

\begin{State}{SouthDakota}
South Dakota does not have an ``approved" statewide document for conducting L\&A, instead relying on its statutes~\cite{south-dakota-admin-rule, south-dakota-election-code}. Communication with state election officials revealed strong functional requirements. Notably, the state ensures proper overvote validation by requiring, for each contest, that election officials mark the maximum number of allowable votes on a test ballot and then exactly one more vote than what is permissible on another.  South Dakota would benefit from requiring election officials to mark test ballots using a source of randomness and by creating publicly-accessible documentation that details statewide requirements. 
\end{State}

\begin{State}{Tennessee}
Tennessee's L\&A policy can be found in state law~\cite{tennessee-code, tennessee-rules}. It allows candidate, news media, and (depending on the type of election) political party representatives to observe testing, but not the general public. Notably, the state earned no credit for ST1, ST2, and ST3 because it only requires testing of tabulators in a number of precincts equal to at least 1\% of the total number of precincts in an election. The state did not responded to requests for additional information. 
\end{State}

\begin{State}{Texas}
Texas is positioned to greatly strengthen its L\&A testing~\cite{texas-regulations} by explicitly requiring the implementation of more rigorous practices. For example, although the state requires testing prior to a given election, it does not require that every device be tested, leaving the devices that are not tested susceptible to preventable errors. The state also requires that every candidate receive at least one vote, but does not ensure that a different number of votes is assigned to each candidate in a contest, thus leaving the possibility of transposed targets untested. Texas does, however, position its policy as a baseline requirement, so  local jurisdictions have latitude to perform more comprehensive testing.
\end{State}

\begin{State}{Utah}
Although Utah's testing requirements are currently only set forth by statute~\cite{utah-code},  the state's election division is developing a best practices guide during the summer of 2022.  Email correspondence revealed that the state already performs well, meeting most of our criteria in both procedural and functional categories. Utah could earn a perfect score by fully validating overvote thresholds and ensuring that some test ballots are marked using a source of randomness.
\end{State}

\begin{State}{Vermont}
Vermont's most recent  L\&A procedures ~\cite{vermont-tabulator-guide} were produced in June 2022 and  require that public testing incorporate all ballot styles and at least one overvote. Notably, the state  requires election officials to randomly fill ten ballots while keeping different vote counts for all candidates. Since the state demonstrated an awareness of the importance of  marking test ballots in a way that makes the outcome less predictable, we awarded partial credit for ND2, even though the policy does not explicitly require use of a random number generator. Vermont's election division did not responded to a request for additional information.
\end{State}

\begin{State}{Virginia}
Virginia's L\&A procedures are outlined in Chapter 4 of its \textit{General Registrar and Electoral Board Handbook}~\cite{virginia-greb-handbook}. The state requires each locality to test all of its voting equipment with each ballot style prior to an election. However, its testing does not ensure that transposed targets are detected nor does it ensure that overvote thresholds are fully validated. The test procedures also do not incorporate nondeterministic elements. In addition to addressing these functional issues, Virginia should consider permitting members of the public who are not representatives of  a candidate or political party to observe testing.  
\end{State}

\begin{State}{Washington}
Washington State meets all our procedural requirements. It tests all machines twice, first at a pretest and then again at a public proceeding~\cite{washington-advisory}. For all elections in Washington, voters are allowed to mark only one option. Proper overvote validation for jurisdictions in this state would therefore look like ensuring every contest has at least one ballot that votes for precisely two options. Incorporating this practice would be a simple way to strengthen the state's functional score.
\end{State}

\begin{State}{WestVirginia}
West Virginia~\cite{west-virginia-code} meets almost all of our procedural criteria. It requires testing of all tabulators prior to an election, permits additional testing in local jurisdictions, and ensures that every ballot style is included in the test deck. The state could strength functional aspects of its requirements by incorporating greater protections offered by BP2 and OP2 as well as by providing nondeterminism. 
\end{State}

\begin{State}{Wisconsin}
Wisconsin maintains a state L\&A policy that provides for the public testing of all machines with all ballot styles before each election~\cite{wisconsin-handbook}. Functionally, the policy requires at least the simple forms of basic protections and overvote protections. There is also a non-binding recommendation to test ``as many vote combinations as possible.'' We recommend strengthening this so that election officials are required to test a different number of votes for each candidate in a contest and to mark some ballots using a source of randomness. The Wisconsin Elections Commission did not responded to a request for additional information. 
\end{State}

\begin{State}{Wyoming}
Wyoming has a strong L\&A testing policy~\cite{wyoming-statute} which, among other provisions, requires that each candidate in a receive a different number of votes. From additional email correspondence with Wyoming's election division, we were able to verify that all tabulators are publicly tested. Wyoming would benefit from consolidating its L\&A requirements into one resource and incorporating a source of randomness for marking some test ballots.
\end{State}

\end{document}